\documentclass[sigconf]{acmart}

\AtBeginDocument{%
  \providecommand\BibTeX{{%
    \normalfont B\kern-0.5em{\scshape i\kern-0.25em b}\kern-0.8em\TeX}}}

\usepackage[capitalize,noabbrev]{cleveref}
\usepackage{multirow}
\usepackage{makecell}
\usepackage{color, colortbl}
\newcommand{\revise}{\textcolor{black}}
	
\definecolor{Gray}{gray}{0.9}
\definecolor{Gray2}{gray}{0.7}
\usepackage{algpseudocode}

\setcopyright{acmcopyright}
\copyrightyear{2022}
\acmYear{2022}
\acmDOI{10.1145/xxx}

\acmConference[]{}{}{}
\acmBooktitle{}
\acmPrice{15.00}
\acmISBN{mmmmmmmmmmmmmm}

\begin{document}

\title{Extracting Replayable Interactions from Videos of Mobile App Usage}

\author{Jieshan Chen}
\authornote{Work done at Apple.}
\affiliation{
    \institution{Australian National University}
    \state{Canberra}
    \country{Australia}
}
\email{jieshan.chen@anu.edu.au}

\author{Amanda Swearngin}
\affiliation{%
  \institution{Apple}
  \city{Seattle}
  \state{WA}
  \country{USA}
}\email{aswearngin@apple.com}

\author{Jason Wu}
\affiliation{%
  \institution{Apple}
  \city{Seattle}
  \state{WA}
  \country{USA}
}\email{jason_wu2@apple.com}

\author{Titus Barik}
\affiliation{%
  \institution{Apple}
  \city{Seattle}
  \state{WA}
  \country{USA}
}
\email{tbarik@apple.com}

\author{Jeffrey Nichols}
\affiliation{%
  \institution{Apple}
  \city{Seattle}
  \state{WA}
  \country{USA}
}
\email{jwnichols@apple.com}

\author{Xiaoyi Zhang}
\affiliation{%
  \institution{Apple}
  \city{Seattle}
  \state{WA}
  \country{USA}
}\email{xiaoyiz@apple.com}

\renewcommand{\shortauthors}{Chen, et al.}

\begin{abstract}
Screen recordings of mobile apps are a popular and readily available way for users to share how they interact with apps, such as in online tutorial videos, user reviews, or as attachments in bug reports. Unfortunately, both people and systems can find it difficult to reproduce touch-driven interactions from video pixel data alone. In this paper, we introduce an approach to extract and replay user interactions in videos of mobile apps, using only pixel information in video frames. To identify interactions, we apply heuristic-based image processing and convolutional deep learning to segment screen recordings, classify the interaction in each segment, and locate the interaction point. To replay interactions on another device, we match elements on app screens using UI element detection. We evaluate the feasibility of our pixel-based approach using two datasets: the Rico mobile app dataset and a new dataset of 64 apps with both iOS and Android versions. On these datasets, we evaluate the performance of our video segmentation, interaction classification, and interaction localization methods with insights. Our approach can successfully replay a majority of interactions (iOS--\revise{84.1\%}, Android--\revise{78.4\%}) on different devices, which is a step towards supporting a variety of scenarios, including automatically annotating interactions in existing videos, automated UI testing, and creating interactive app tutorials.

\end{abstract}


\maketitle

\section{Introduction}
\label{sec:intro}

Videos of mobile app usage have become commonplace on the internet. For example, Tech vloggers make app tutorials to educate new users or share the best practices of app features. People record app screens step-by-step to guide their parents who are not familiar with an app, and app testers create bug reports with rich context in video.
In particular, screen recordings---that is, videos that record on-screen content but not the user's hands---are easy to produce and share using built-in smartphone facilities.

In order to effectively use these screen recordings however, the viewer has to first understand the interactions performed in the video and then manually repeat them in the order shown. This process can be time-consuming and error-prone~\cite{bernal2020translating}, especially when the sequence of necessary interactions is long or the recording is played quickly. For example, a viewer may have to pause the recording after each interaction, or even replay it multiple times before they are able to replicate the interaction on their own device. In addition, the target UI element for an interaction may be difficult to locate in complex app screens, in the presence of different display preferences---such as large fonts or dark mode---or when device or app versions change.

What if instead of asking a user to do all of the work of interpreting a screen recording, a machine could do it instead? Such a system might identify the type of interactions that were performed and the target UI elements that these interactions were performed on, ideally from the recording alone. Work has been attempted in this area before, but a limitation of existing systems is that they require a special recording apparatus~\cite{yu2019lirat, li2018appinite}, visual indicators added during recording~\cite{qian2020roscript, bernal2020translating} or additional metadata such as UI transition graph~\cite{feng2022gifdroid} in order to capture the interactions demonstrated in the video.

In this paper, we propose a system that automatically extracts user interactions from ordinary screen recordings, without requiring additional settings, specially-instrumented recording tools, or source code access. Our system performs three phrases to extract the interactions: 1) video segmentation, 2) interaction classification, and 3) interaction localization.
Our system first segments the start and the end of each interaction. Then, it runs heuristics to choose between six common interaction types. Finally, our system uses a 3D convolutional encoder to learn the semantics of the animation in the UI, a 2D convolutional encoder and decoder to capture the connections between consecutive UI states, and another decoder to infer the interaction probability heatmap and output the location of the interaction. Based on UI detection results from screenshot pixels, we can find the target UI element and its content. We also explore methods to replay the interactions that we extract from videos on another device. Our replay prototype runs UI detection on each app screen, locates the best matched UI element for each recorded interaction and performs the interaction on each UI element in turn.
Table~\ref{tab:technique_differences} summarizes the key differences between our method and existing techniques.

We evaluated our system on the Rico dataset \cite{deka2017rico} (created 4 years ago), and a smaller app usage dataset (64 top-downloaded apps, each has iOS and Android versions) that we collected and annotated recently. For video segmentation, our system achieves \revise{84.7\%} recall on iOS, and slightly worse recall on Android (\revise{72.0\%}). For interaction classification, our system achieves comparable accuracy on both platforms (iOS--\revise{87.6\%}, Android--\revise{89.3\%}). For interaction localization, our system achieves the best accuracy on Rico (\revise{69.1\%}), as the model is trained on this dataset. Although app UI design changes have occurred since Rico, our model still works on recent Android with lower accuracy (\revise{56.2\%}), and \revise{41.4\%} accuracy on recent iOS apps. For interaction replay, we found that the majority of interactions (iOS--84.1\%, Android--78.4\%) can be replayed on different devices.

The contributions of this paper are as follows:
\begin{itemize}
    \item We present a pixel-based approach to automatically extract and replay interactions from ordinary screen recordings without requiring additional settings, specialized recording instrumentation, or access to source code.
    \item We implement a prototype system that instantiates our approach. The results of our evaluation show reasonable accuracy in video segmentation, interaction classification, and interaction localization. Our system successfully replays a majority of interactions on different devices. These results demonstrate the feasibility of our pixel-based approach to extracting replayable interactions.
\end{itemize}

\begin{table} [b]
  \caption{Differences in input, additional data requirements, and support for cross-device replay between existing techniques and our proposed system.}
  \label{tab:technique_differences}
  \resizebox{0.5\textwidth}{!}{
  \begin{tabular}{llllc}
    \toprule
                              & Input &  \makecell[l]{Additional\\ Requirement} & \makecell{Cross-Device \\Replay} \\
    \midrule
    \rowcolor{Gray}
    \texttt{APPINITE\cite{li2018appinite}}     & User demonstration  & \makecell[l]{View hierarchy, { } { } { }  { } \\Interaction\\point \& type}  & No \\ 
    \texttt{V2S\cite{bernal2020translating}}   & Video                     & \makecell[l]{Add touch indicator\\to the video}           & No  \\
    \rowcolor{Gray}
    \texttt{RoScript\cite{qian2020roscript}}   & \makecell[l]{Test script or Video}       & Include fingertips            & No      \\
    \texttt{LIRAT\cite{yu2019lirat}}           & \makecell[l]{Test script or\\User demonstration} & \makecell[l]{Interaction \\point \& type}  & Yes  \\
    \rowcolor{Gray}
    \texttt{GifDroid\cite{feng2022gifdroid}}   & \makecell[l]{Video and UI { } { } { } { }\\ Transition Graph { } { } }       & None            & No      \\
    \rowcolor{Gray2}
    \hline
    \texttt{Our system}                         & \textbf{Video}                     & \textbf{None}                  & \textbf{Yes} \\
    \bottomrule
  \end{tabular}
  }
\end{table}

\section{Related Work}
We discuss the related work across two areas: 1) identifying interactions on user interfaces, and 2) replaying user interactions.

\subsection{Identifying Interactions on User Interfaces}
Identifying user interactions is an important task on various platforms, including mobile~\cite{li2018appinite,bernal2020translating,yu2019lirat,qian2020roscript, feng2022gifdroid}, desktop~\cite{zhao2019actionnet, nguyen2015making}, web~\cite{bao2017extracting, sprenkle2005automated}, and even the physical world~\cite{guo2019statelens}. The extracted interactions can empower many applications, including task automation~\cite{li2018appinite}, bug reporting~\cite{bernal2020translating}, automated app testing~\cite{yu2019lirat,qian2020roscript, sprenkle2005automated}, and guidance to use appliances~\cite{guo2019statelens}.

Previous research has applied various methods to identify user interactions for the purposes of replaying them on the same or another device. LIRAT~\cite{yu2019lirat} obtains the interaction location and type by using a debugging tool to access low-level system events. The device must also have a connection to a computer that runs the debugging tool. APPINITE~\cite{li2018appinite} adds a layer of interaction proxies~\cite{zhang2017interaction} on top of the current running app. An interaction proxy layer captures the users' taps and passes them to the underlying app. This method requires installing an additional background service and obtaining Accessibility permissions, and may not work on all platforms. V2S~\cite{bernal2020translating} requires users to access Android developer settings in order to show a touch indicator at each \textit{tap} event. With this known visual indicator, V2S presents an object detection model to locate the touch indicator and infer the interaction location. This method adds extra work to app video creators, and not all video creators would like to show a developer-mode touch indicator in videos.
RoScript~\cite{qian2020roscript} instead requires video creators to use an external camera to record the phone screen and finger movement. It leverages computer vision techniques to recognize a finger and its relative location to the app UI, and the system requires users to move their fingers outside the phone screen between each interaction for segmentation. This method also requires an additional camera and a stable setup of phone and camera. While GifDroid~\cite{feng2022gifdroid} does not require complex setup of the input video, it additionally relies on the UI transition graph to assist interaction identification. However, constructing UI transition graph is not be a trivial graph and requires many efforts.

The methods above require settings or recording tools that are specific to a platform (e.g., Android) \cite{yu2019lirat, li2018appinite}, add extra work to app video creators \cite{bernal2020translating, qian2020roscript}, and will not work on existing app usage videos \cite{yu2019lirat, li2018appinite, bernal2020translating, qian2020roscript}. We believe our pixel-based approach can be a more generalizable way to collect interaction traces from videos.

\subsection{Replaying User Interactions}

After extracting user interactions, the key challenge of replaying is to find where to interact on the replay device. Some work repeats the (x, y) coordinate from the recording~\cite{bernal2020translating, halpern2015mosaic}, while some applications~\cite{yu2019lirat, qian2020roscript} find matching UI elements on replaying devices so that the replay will be more robust to dynamic content and device change.

To find a matching UI element sometimes requires access to a view hierarchy. For example, APPINITE~\cite{li2018appinite} tries to match UI metadata from the view hierarchy (e.g., parent-child relationship, text, UI Class) so that it can still locate the target UI element even when the target UI element moves to a different location due to screen content change. However, the view hierarchy is not always available, and the view hierarchy can be incomplete or misleading.

To avoid these limitations, some work leverage computer vision techniques to match targeted UI elements.
For example, LIRAT~\cite{yu2019lirat} compares image features to match UI elements between recording and replaying screens, and extracted layout hierarchy from pixels to improve matching. Similarly, our method also leverages video pixels to match target UI elements, but we use object detection models that have better performance.

\begin{figure*}
	\centering
	\includegraphics[width=1\textwidth]{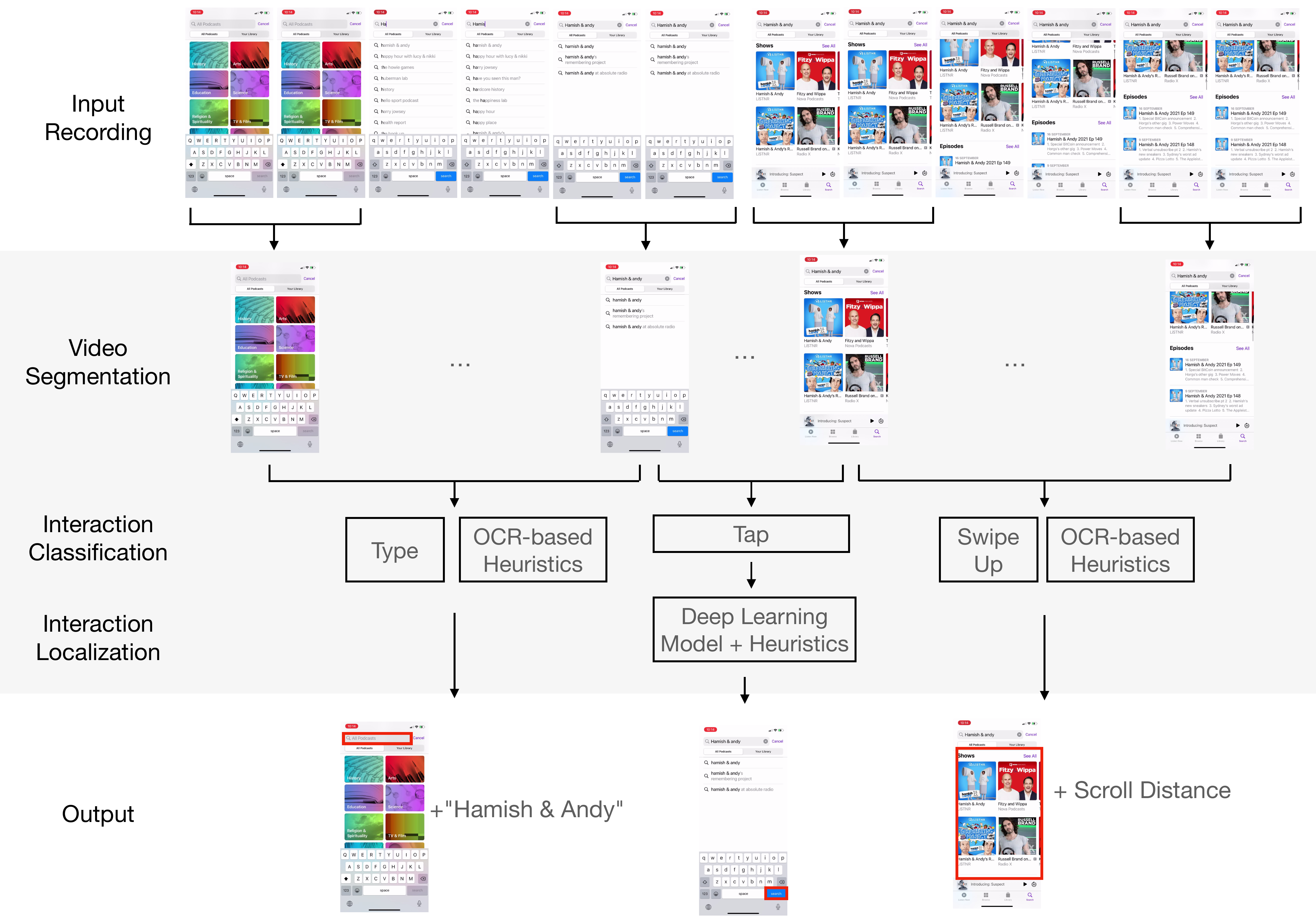}
	\caption{Flowchart of our system: First, it segments the frames from an input screen recording into a sequence of representative frames (i.e., keyframes). Second, it applies heuristics to classify the interactions into six common types. For \textit{type} and \textit{swipe} interactions, it relies on the OCR heuristics to locate the interaction point and the corresponding information (typed text for \textit{type} interactions, scroll distance for \textit{swipe} interactions). For \textit{tap} interactions, it applies both heuristics and a deep learning based interaction localization model to determine the interaction point. The overall output is an interaction trace that can be replayed on another device. }
	\label{fig:flowchart}
	\Description{The flowchart is partitioned into 5 rows. The first shows several screenshot frames from an input recording with arrows below sections corresponding to keyframes pointing to the second row.  The second row is labeled ``Video Segmentation'' and shows a reduced set of screenshots indicating keyframes. The third row shows the keyframes from the second row being classified into one of several interaction types, including \textit{type}, \textit{tap}, and \textit{swipe}. Once the interaction types have been determined, the outputs of the third row have arrows going into associated OCR-based heuristics, or localization models for each interaction type. Finally, the last row, Output, shows a screenshot and inputs needed to generate an interaction trace for each type. For example, the \textit{type} interaction shows an arrow going to a box labeled OCR-based heuristics, and the final output is a screenshot with a box around the detected Type UI element, and the text ``Hamish & Andy''. }
\end{figure*}

\section{System}

Essentially, our system extracts interactions from pixels in video frames, and uses this information to replay the interactions on another device. \revise{As shown in Figure~\ref{fig:flowchart},} extraction is performed through three phases: 1) video segmentation (\cref{sec:video_segmentation}), interaction classification (\cref{sec:interaction_classification}), and interaction localization (\cref{sec:interaction_localization}). Our system then applies a set of strategies in the interaction replay phase (\cref{sec:interaction_replay}). The rest of this section describes the system phases in detail.

\begin{figure*}
	\centering
	\includegraphics[width=1\textwidth]{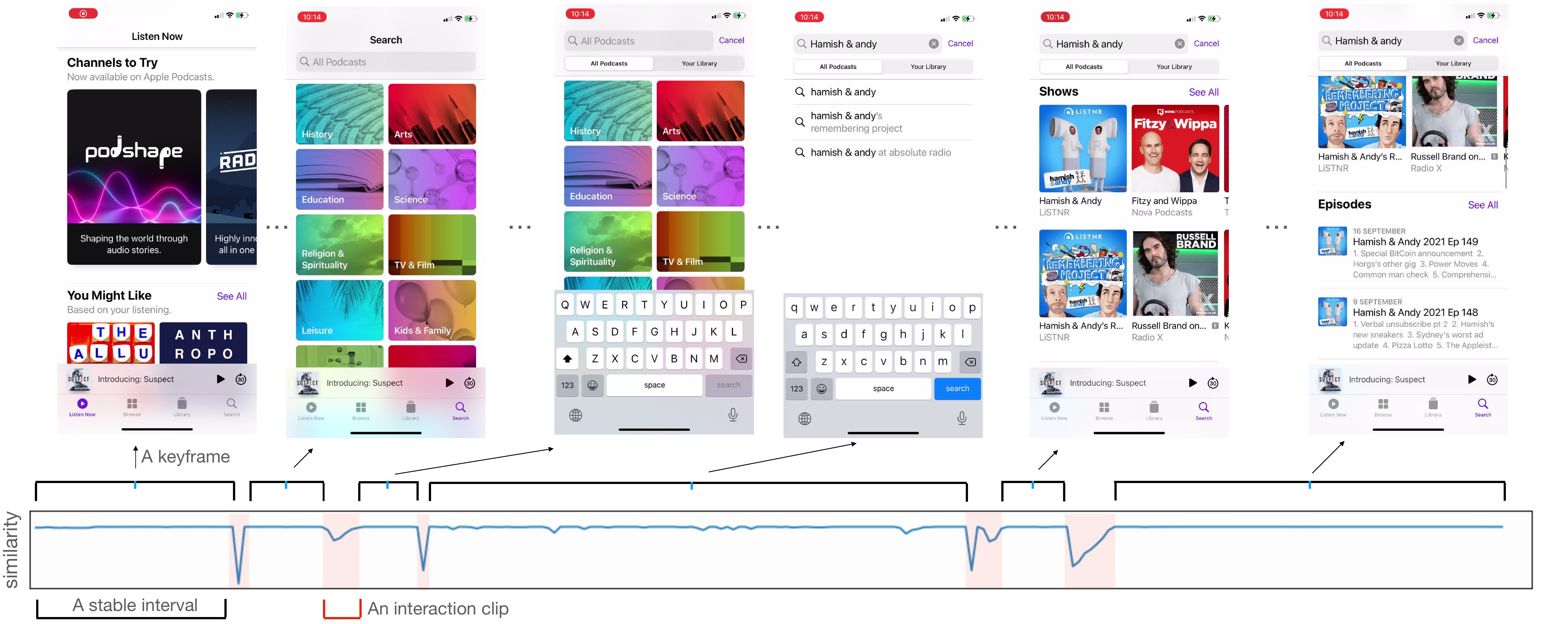}
	\caption{Visualization of image similarities between consecutive frames. The spikes indicate segments in the video where user interactions were performed, which we use to segment the keyframes. In this figure, we detected six stable intervals, and for each interval, we take the middle frame as the extracted keyframe. \revise{The users \textit{tap} on the bottom-right ``search'' icon, \textit{tap} the input field on the top, \textit{type} text, \textit{tap} the bottom-right ``search'' icon, and \textit{swipe} up to see more content.}
	}
	\label{fig:segmentation}
	\Description{The top of the figure shows several screenshots from an interaction trace split up into keyframes. The bottom of the figure shows a line graph of structural similarity over time, as the interaction trace proceeds. The line graph shows several "spikes" where the structural similarity value goes drastically down as the user interactions are performed, and flat areas where the structural similarity is staying the same corresponding to the stable intervals.}
\end{figure*}

\begin{figure*}
	\centering
	\includegraphics[width=0.8\textwidth]{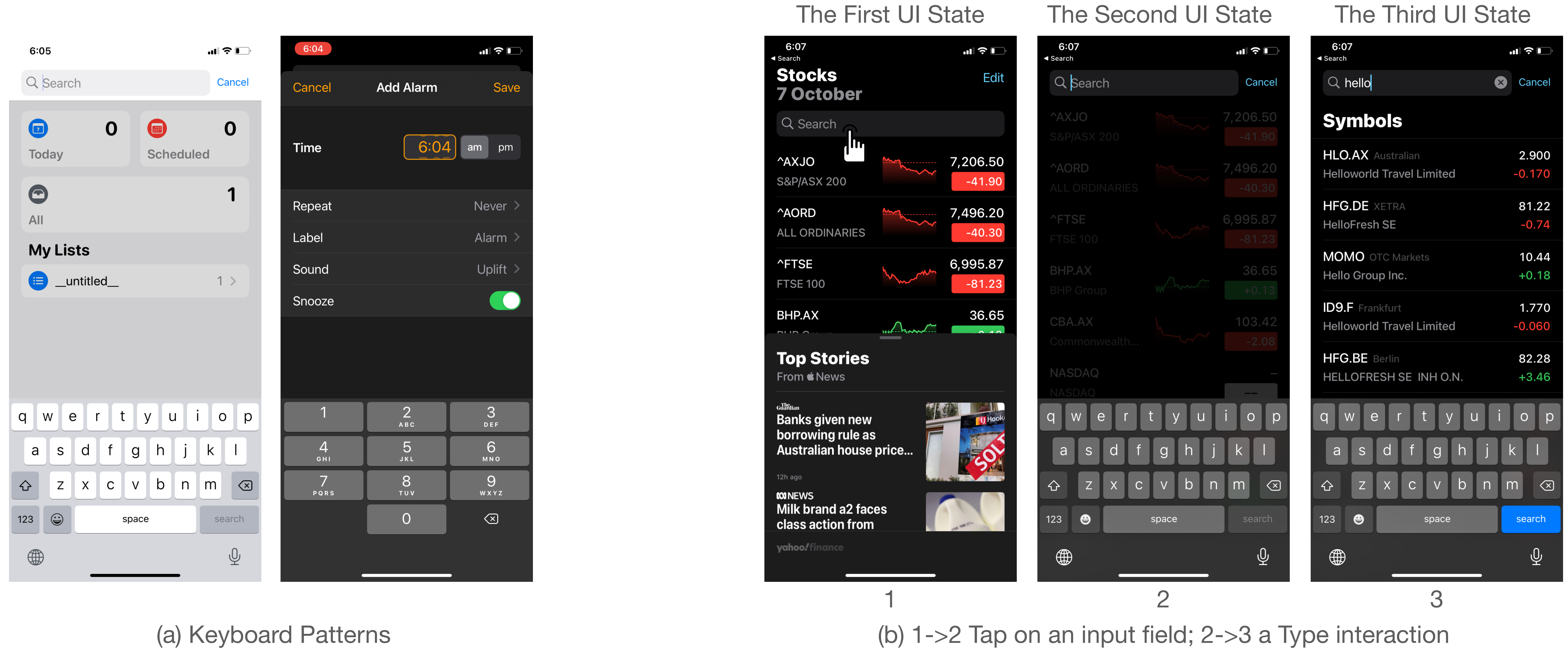}
	\caption{Examples of patterns that our interaction classification heuristics examine to classify a \textit{type} interaction, including (a) Keyboard patterns that can be detected from OCR results. (b) When users \textit{tap} on an input field, the title of the UI will change instantly; when users perform a \textit{type} interaction, the title will have a steady change or remain the same.
	}
	\Description{On the left, we show two screenshots of an app screen with keyboards open, one on the QWERTY keyboard and one on a number keyboard to show how we distinguish a typing action. On the right, we show three screenshots. The first is a Stocks page with a search field on the top, the second screen shows the search field active with the keyboard open, and the third screen shows the same screen where some text has been typed in a search field. The indicated interaction is that the user tapped on the search field to activate the keyboard, and typed some text.}
	\label{fig:type_action}
\end{figure*}

\begin{figure*}
	\centering
	\includegraphics[width=0.8\textwidth]{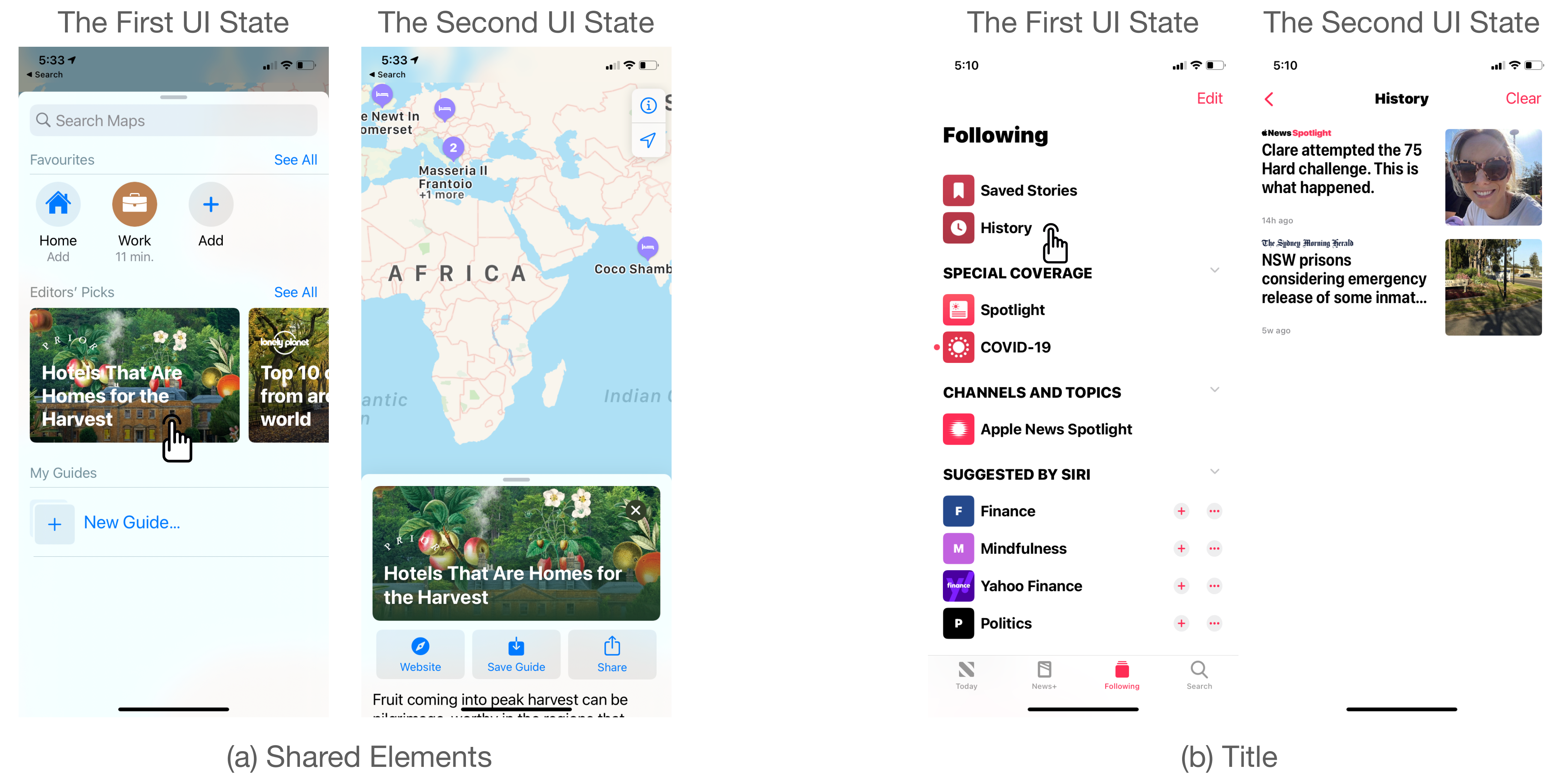}
	\caption{Examples of patterns that our interaction localization heuristics look for to localize a \textit{tap} interaction, including (a) showing the start and the end UI state when the user taps on an item, and the item remains in the next page having the same label and the same image, and (b) showing an example when the tapped item's label becomes the title of the next UI state.}
	\label{fig:localization_heuristics}
	\Description{The figure shows two examples of heuristics our system applies for interaction localization. On the left, we show two UI screens in a Maps app that share some UI elements, including an image with text (which the user tapped) which was moved down in the second UI screen below a map indicating the user tapped on it. On the right, we show an example of a News app UI screen. The first UI screen shows a list of items, one labeled "History". The second UI screen has the title "History" indicating that the "History" list item was tapped.}
\end{figure*}

\subsection{Phase 1---Video Segmentation}
\label{sec:video_segmentation}

Video segmentation splits the frames of the input screen recording into a sequence of representative frames that maximally differentiates the video, or \emph{keyframes}~\cite{zhong1996clustering}.

To illustrate how this works, consider the frames of the screen recording shown in \cref{fig:segmentation}. We start by computing the histogram of oriented gradient (HOG)~\cite{dalal2005histograms} feature descriptor for each frame. As the name suggests, the HOG descriptor is a simplified representation of the screen in terms of its structure or shape through gradient and orientation. We use this HOG descriptor to calculate similarity between consequence frames using a structural similarity (SSIM) measure~\cite{wang2004image}. Intuitively, a sequence of similar frames represents a stable interval---with the middle of this stable interval being the keyframe, represented with an arrow in \cref{fig:segmentation}.

Next, we run a spike detection algorithm using empirical parameters we derived from a number of app usage videos (\cref{sec:evaluation}): 1) the spike should be larger than $(\mathrm{Similarities}_{\mathrm{max}} - \mathrm{Similarities}_{\mathrm{min}})/15$~\footnote{\revise{15 is an empirically set number from experiments.}}, to be resilient to partial content changes that are not caused by user interactions, and 2) the stable intervals should contain at least four frames, to mitigate against interactions from transient UI changes. The inverted spikes in \cref{fig:segmentation} indicate interaction clip points that segment the keyframes.

\subsection{Phase 2---Interaction Classification}
\label{sec:interaction_classification}
This phase identifies six interactions---\textit{type}, \textit{swipe left}, \textit{swipe right}, \textit{swipe up}, \textit{swipe down}, and \textit{tap}~\cite{li2019humanoid}---through the following heuristics:

\textbf{\textit{Type}} interactions are always associated with a virtual keyboard. We inspect the OCR results on a screen to determine if they contain text corresponding to the rows of a keyboard---that is, 3 rows of QWERTY keyboard, and 4 rows of number pad (Figure~\ref{fig:type_action}(a)). For entered text, we compare the OCR results in the first and last frames of a \textit{type} interaction. To illustrate, in Figure~\ref{fig:type_action}(b)'s right-most screen, the placeholder inside the top search bar changes as text is entered, with suggestions appearing below. Among all changed or added OCR text results in the last frame, we pick the top-most one (smallest $y$) as the entered text.

\textbf{\textit{Swipe (left, right, up, down)}} interactions shift several UI elements within a scrollable area, while the top title bar and bottom tab bar are often unchanged (see \cref{fig:swipe_left_vs_tap} and \cref{fig:swipe_up_vs_tap} in the Appendix). Consequently, we compare the OCR results between any two consecutive frames and calculate the movement between each pair of text strings. If multiple ($N>=3$, an empirical parameter) text strings move in the same horizontal or vertical direction within a threshold distance, our system classifies the interaction as a \textit{swipe}. We call the text strings with shared movement a \emph{text collection}. \revise{Note that sometimes ``snackbar'' elements may briefly appear at the bottom of the screen with messages about app processes, so we set $N>=3$ to avoid confusing this UI behavior with a swipe.}

Capturing the semantics of \textit{swipe} interaction requires three properties: the direction, the distance, and initiation point. The Swipe direction is determined trivially through the movement coordinates. To calculate the \textit{swipe} distance, we use the median distance of movement between two consecutive frames, and then sum all median distances between any two consecutive frames in the interaction clip. The \textit{swipe} initiation point is the either first or last text by x or y position, for horizontal or vertical \textit{swipe}, respectively. 

\textbf{\textit{Tap}} interactions may lead to a new UI state, pop-up keyboard, or cause few or no movement of shared elements. If an interaction clip is not classified as either a \textit{type} or \textit{swipe}, we classify it as a a \textit{tap} interaction. From the Rico dataset~\cite{deka2017rico}, we found that the majority of interactions in mobile apps are \textit{tap} interactions (91.7\%), and that \textit{type} and \textit{swipe} interactions often cause changes on text elements---for example, through creating or moving text. Informed by these findings, treating \textit{tap} as the fall-through interaction has reasonable justification.

For a \textit{tap} interaction, we must also identify the \textit{tap} location. This is described in the next section.

\subsection{Phase 3---Interaction Localization for Tap Interactions}
\label{sec:interaction_localization}

For \textit{tap} interactions, interaction localization is needed to identify the location of the \textit{tap}. In some cases, the start and end UI state will share an interaction component. For this situation, we can use heuristic-based localization (\cref{sec:heuristic-based_localization}) to identify the \textit{tap} interaction location.

When heuristic-based location fails, we can rely on visual feedback cues provided by the app when the users \textit{tap} a location. In this situation, we leverage the animation effect and the connections between the two consecutive UI states to train an interaction localization model to locate the interaction point (\cref{sec:localization_model}).

\begin{figure*}
	\centering
	\includegraphics[width=1\textwidth]{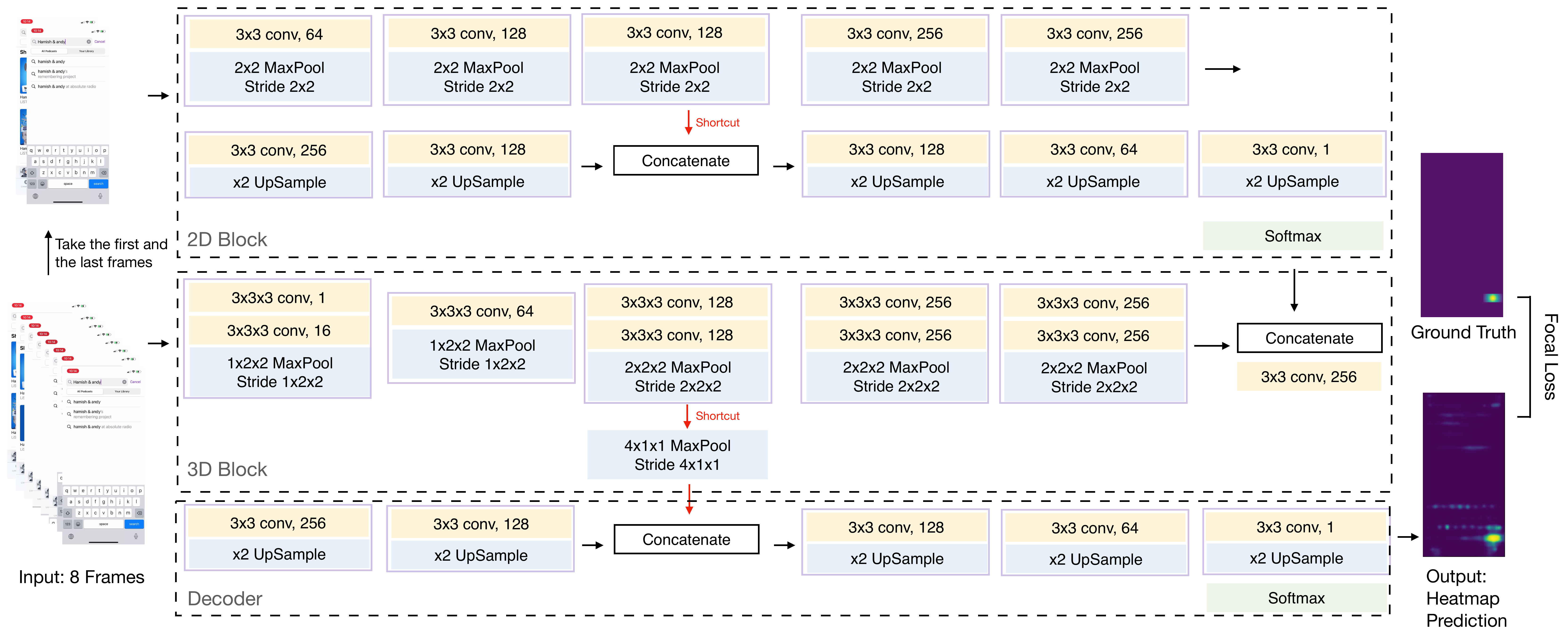}
	\caption{The structure of our interaction localization model. Given eight frames as the input, the model takes the first and the last frames as the input to the 2D block to learn the connections between the two consecutive UI states; Concurrently, the model feeds all eight frames into a 3D block to encode the animation effect; the model later concatenates the extracted features from the 2D block with the animation features extracted from the 3D encoder, and the combined features are then feed into a decoder to predict the interaction heatmap.
	}
	\Description{The figure shows the structure of the proposed localization model. From left to right, it consists of the input frames, the detailed parameters of our model structure and the output heatmap.For the leftmost input, it shows a stack of 8 frames extracted from an interaction clip, which is the input of our 3D block module, and we takes the first and the last frame as the input of the 2D block module.For the middle model structure, from top to down, it consists of three sub-blocks, namely a 2D block, a 3D block and a decoder.The rightmost shows the output, i.e., the predicted interaction heatmap, as well as the ground truth heatmap. We use a variant of focal loss as our target loss function.}
	\label{fig:localization_model}
\end{figure*}

\subsubsection{Heuristic-based Localization}
\label{sec:heuristic-based_localization}

When the title of the new UI state is same as the label of one items in the content area, it is likely that this is the item the user \textit{tap}ped. In Figure~\ref{fig:localization_heuristics}(b), when users \textit{tap} on ``History'', the title of the new UI state also becomes ``History''. Because this is a high-accurate heuristic, we first detect the existence of this pattern to locate the interacted element (\cref{sec:heuristic-based_localization}). 

To detect the title, we first run OCR on the first and last frame of each interaction clip to obtain all texts. We then find the top title in the second frame, and check if there is an element with the same text in the main content---excluding the top bar and the bottom app bar. If so, we output the position of this element as current \textit{tap} interaction point.

\subsubsection{Localization Model}
\label{sec:localization_model}

From our observations of the Rico animation data and our collected app usage videos, we identified three common visual cues that we can leverage to locate \textit{tap} interactions: 1) ripple effect---a radial action in the form of a visual ripple expanding outward from the user's touch, 2) expand effect---which scales up and cross-fades a UI elements, and 3) changes in the text or background colors.
In addition, we noticed that in some cases, the start and end UI state shares the interacted element. For example, in Figure~\ref{fig:localization_heuristics}(a), when users \textit{tap} ``Hotels That Are Homes for the Harvest'', the new UI state contains the same text. However, although in this case, the shared element is the interacted element, it may fail in other situation. We rely on our models to learn the difference between normal shared elements and shared interacted elements.

We trained a deep-learning based interaction localization model to locate the \textit{tap} point. Inspired by the success of human pose recognition~\cite{pfister2015flowing}, object detection~\cite{law2018cornernet, lin2017focal}, and video classification models~\cite{tran2015learning}, we designed a model to predict a heatmap of possible \textit{tap} points by learning the semantics of the animation effects.

As shown in Figure~\ref{fig:localization_model}, our model primarily consists of three blocks: a 2D block, a 3D block and a decoder.
The 2D block is based on 2D convolutional layers~\cite{pfister2015flowing} and aims to find the connections between two consecutive UI states, while the 3D block---based on 3D convolutional layers~\cite{tran2015learning}---captures the temporal relationship among frames---that is, the animation effect across these frames---in each interaction.
A final decoder then fuses features extracted from the 2D and the 3D block to infer the interaction heatmap.
We added a shortcut module following the U-Net model~\cite{ronneberger2015u} to help the model to retrieve the coarse features from the shallow layer in the encoder part to refine the extracted high-level abstract features and help dense per-pixel heatmap prediction. Concretely, given 8 frames extracted from one interaction clip as the input of our model (\cref{sec:evaluation:interaction_localization}), we take the first and the last frames as the input to 2D block to learn the connections between the two consecutive UI states. In parallel, we feed all frames into our 3D block to encode the animation effect: the extracted features from 2D block and 3D block will then be concatenated, and the combined feature will be fed into the decoder to predict the interaction heatmap, which is the output of our localization model. We take the point with highest probability in the predicted heatmap as the output interaction point.

Given that we only have one \textit{tap} point in each training sample, there will be only one out of all 256x512 points in the heatmap being set to 1, while all other points are set to 0. Therefore, our dataset has a similar data imbalance problem as encountered in many object detection tasks~\cite{lin2017focal}. We used two strategies to alleviate this issue. First, instead of setting only one point in the heatmap as 1 and others as 0, we found the bounding box of target UI element (\cref{sec:interaction_clips_from_rico_dataset}) and then applied the 2D Gaussian function to obtain the probability of surrounding points in the interaction elements~\cite{law2018cornernet, pfister2015flowing, duan2019centernet}.
Second, we used a variant of the focal loss~\cite{lin2017focal, law2018cornernet} to perform a weighted penalization on the low confidence data: let $p_{ij}$ be the predicted score (a.k.a. confidence) at location (i,j) in the predicted heatmap, and let $y_{ij}$ be the ground-truth score augmented by the 2D Gaussians. 
Then, the loss function is: 
$$
L = -\frac{1}{N} \sum_{i=0}^{H} \sum_{j=0}^{W} \left\{
        \begin{aligned}
            (1-p_{ij})^{\alpha}log(p_{ij}) & \;                    & if \; y_{ij} = 1 \\
            (1-y_{ij})^{\beta}(p_{ij})^{\alpha}log(1-p_{ij}) & \;  & otherwise
        \end{aligned}
    \right.
$$

We used the interaction trace and animation datasets in Rico dataset~\cite{deka2017rico} to train our model. The details are explained in \cref{sec:interaction_clips_from_rico_dataset}.
Our localization model is trained on 4 Tesla V100 GPUs using an Adam optimizer~\cite{kingma2014adam} for 25 epochs with the initial training rate being 0.01 and a batch size of 32. For each interaction clip, we evenly picked 8 frames from the interaction clips as model input. We can also pick more frames as input, and we evaluated its impact in \cref{sec:evaluation:interaction_localization}
Therefore, the structures of 3D blocks for these two inputs are slightly different, with the 3D pooling layer in the third convolutional block having a depth stride of 1 or 2.
If the interaction clip does not contain 8 frames, we duplicate some of the frames to get 8 frames.
During training, the first and the last frame is fed into the 2D block and extract features from the first and the next UI state.
In parallel, all frames are fed into the 3D block to extract the semantics of the animation effect.
The two extracted features from 2D and 3D blocks are then concatenated and fed into a decoder to predict the interaction heatmap. The original size of Rico animation clips are 281x500 (width x height), and we resized them to 256x512.

\subsection{Phase 4---Interaction Replay}
\label{sec:interaction_replay}

The previous phases extracted interactions from video. When replaying interactions on another device, we sometimes can directly repeat interactions on screen (for example, typing entered text), but otherwise need to find a matching target UI elements to apply the interactions---such as a UI element to tap or a point to start swiping.

To accomplish interaction replays, we run an object detection model~\cite{zhang2021screen} to detect UI elements on each keyframe of recorded video, and then find the UI detection that contains a \textit{tap} point or a \textit{swipe} initiation point; if there are multiple detections, we pick the smallest detection. To find the target UI on the screen of a new device, we run fuzzy matching~\cite{fuzzywuzzy} for text elements and leverage template matching~\cite{matchTemplate} for non-text elements.

Specifically, we choose the text with the highest weighted ratio (case-insensitive, ignore punctuation)~\cite{fuzzywuzzy} as the matching target text element. For non-text element, we use it as the template image, and slide it over the new screenshot to find a location with the highest matching value. We used normalized correlation coefficient as our matching function. When the replaying devices have different resolutions than the recorded video, we scale (50\% to 200\%) accordingly to the template image so that its size is similar to the target UI element in new screenshot---that is, multi-scale template matching.

Running matching algorithms on every pixel of screenshot can be time-consuming. To speed up matching, we first limit the search space to the detected UIs on the new screen, and run matching on full screenshot only when we fail to find a match from UI detections.
\section{Evaluation}
\label{sec:evaluation}

We evaluated our system on a large-scale dataset (Rico~\cite{deka2017rico}, created 4 years ago), and a smaller app usage recording dataset (iOS and Android versions of 64 top-downloaded apps) we collected and annotated recently (\cref{sec:evaluation:datasets}). We evaluated each phase of system: video segmentation (\cref{sec:evaluation:video_segmentation}, iOS--84.7\%, Android--72.0\% recall), interaction classification (\cref{sec:evaluation:interaction_classification}, iOS--87.6\%, Android--89.3\% accuracy), interaction localization (\cref{sec:evaluation:interaction_classification}, Rico--69.1\%, Android--56.2\%, iOS--41.4\% accuracy), interaction replay across devices (\cref{sec:evaluation:interaction_replay}, iOS--84.1\%, Android--\revise{78.4\%} success rate).

\subsection{Datasets}
\label{sec:evaluation:datasets}

\subsubsection{Interaction Clips from Rico Dataset}
\label{sec:interaction_clips_from_rico_dataset}

The Rico dataset~\cite{deka2017rico} is a large-scale repository of Android app screens. In addition to UI element information on each screen (e.g., bounding box, UI class), the dataset also contains interaction traces of the apps and their corresponding video clips---for example, displaying animations after performing each interaction.

Each interaction trace provides a list of gestures to perform interaction, and we needed to derive an interaction type and interaction point from each gesture. We consider six interaction types as in ~\cite{li2019humanoid}, namely \textit{tap}, \textit{swipe left}, \textit{swipe right}, \textit{swipe up}, \textit{swipe down} and \textit{type}.
We adopted the following heuristics from ~\cite{li2019humanoid} to determine \textit{tap} and \textit{swipe} interactions:

\begin{enumerate}
    \item If an interaction contained only one gesture point or the distance of the gesture was $\leq$ 10 pixels, we considered it a \textit{tap} interaction.
    \item If an interaction contained a list of gesture points with distance $>$ 10 pixels, we considered it a \textit{swipe} interaction. We mapped the gesture direction to \textit{swipe left}, \textit{swipe right}, \textit{swipe up}, or \textit{swipe down}.
\end{enumerate}

For the \textit{type} interaction, we noted that Rico dataset workers used physical keyboards to type text, and the \textit{type} interactions were not recorded in the gesture data as a result. From our observations, we noted that the \textit{type} interactions happened after a \textit{tap} interaction on text field. Therefore, we detected these \textit{tap} interactions and text changes in the text field, and then manually verified these potential \textit{type} interactions.

In total, we obtained 44,536 interactions (with interaction type and video clip) from 7,211 user interaction traces in 6,547 free Android apps. Among these interactions, 91.7\% (40,855/44,536) are \textit{tap}, 0.3\% (123/44,536) are \textit{type}, 5.2\% (2,299/44,536) are \textit{swipe up}, 1.0\% (442/44,536) are \textit{swipe down}, 1.54\% (688/44,536) are \textit{swipe left}, and 0.3\% (129/44,536) are \textit{swipe right}.
We found several limitations in Rico interaction traces and their video clips. Because the dataset only contains clips---and not a continuous usage recording---we were unable to evaluate our segmentation method using Rico. Some interactions omit their video clip or gesture, while other gestures do not match the video clips. Some video clips contain no changes in the UI. These data quality issues may significantly impact interaction classification result (especially on less frequent classes), as the interaction types are already highly skewed. Nevertheless, we were able to evaluate Rico on our interaction localization model---as our identified data issues have negligible impact for \textit{tap} interactions. Thus, we used the Rico \textit{tap} interactions to train our interaction localization model, and report our model performance on this testing split. \revise{Note that as we only use \textit{tap} data to train the localization model, so that the localization model will not be biased.}

An additional limitation of the Rico dataset is that it contains only Android apps, and was collected four years ago. As a result, this dataset may not reflect recent app designs on major mobile platforms. Thus, we collected and annotated usage recordings from top-downloaded iOS and Android apps---as discussed in the next section.

\begin{table*}
  \caption{The total number of interactions for each interaction type, and average task duration across 128 collected recordings from 64 top-downloaded applications.}
  \label{tab:collected_recordings}
  \begin{tabular}{lrrrrrrrr}
    \toprule
                        &   \#{Taps} &  \#{Types} & \#{Swipe-Ups} &  \#{Swipe-Downs} &  \#{Swipe-Lefts} &  \#{Swipe-Rights} & \#{Total} & Avg. Duration  \\
    \midrule
    \texttt{Android}    & 396 & 33  & 78  & 15 & 6 & 6 & 534    & 35.8s  \\
    \texttt{iOS}        & 391 & 36  & 74  & 12 & 3 & 2 & 518    & 29.3s  \\
    \hline 
    \texttt{Total}      & 787 & 69  & 152 & 27 & 9 & 8 & 1,052  & 32.6s \\
    \bottomrule
  \end{tabular}
\end{table*}

\subsubsection{Usage Recordings from Top-Downloaded iOS and Android Apps}

\revise{We followed the process of \citet{bernal2020translating} to collect our dataset, and ensure its diversity and representiveness.}
We collected 64 top-downloaded free apps from the 32 categories in the Australia Google Play store (two apps per app category), all of which also offered a free iOS version to enable fair evaluation between different platforms. 
\revise{To ensure the collected recordings were representative, all authors discussed and selected the tasks, which include the key features of each app based on their description in the both app stores.}\footnote{The task details can be found in our supplementary materials.}
The first and second authors, \revise{one female and one male, both without any disabilities,} randomly picked an app, installed it on both an iPhone 11 (iOS 14, physical device, \revise{1792 x 828}) and Nexus 6P (Android 11, emulator, \revise{2560 x 1440}), and recorded the screen while performing the same task in both the iOS and Android apps. 
\revise{When recording the videos, the two authors used the mobile apps as normal with no restrictions on their interactions.}

Once app usage videos were recorded, the first author manually annotated them to segment stable intervals, classify interaction types, and locate interaction elements.
We used an open-source tool, labelImg~\cite{labelImg}, to facilitate the annotation process.
In total, we obtained 128 app usage recordings ($\mu$ duration = 32.6 seconds), containing 1,052 interactions (787 \textit{taps}, 69 \textit{types}, 196 \textit{swipes}). Additional details are found in \cref{tab:collected_recordings}).

\subsection{Phase 1---Video Segmentation}
\label{sec:evaluation:video_segmentation}

We evaluated our model's performance in video segmentation on usage recordings from top-downloaded iOS and Android apps. We examined each keyframe predicted by our model with all stable intervals in annotated ground truth. We classified our video segmentation are correct when a predicted keyframe falls into a stable interval (with no other predicted keyframes in this interval). We counted the \# of correctly predicted keyframes (C), \# of predicted keyframes (P), and \# of annotated ground truth keyframes (A), and then calculated precision ($ \frac{C}{P}$), recall ($ \frac{C}{A}$) and F1-score.

\begin{table}[t]
	\centering
	\caption{Experimental results for video segmentation on recordings from top-downloaded apps on iOS and Android. We report Precision (P), Recall (R), and F1 score for each combination of feature extraction method and distance function.}
	\begin{tabular}{| l | c c c | c c c |}
		
		\hline
		& \multicolumn{3}{c|}{\textbf{Android}} & \multicolumn{3}{c|}{\textbf{iOS}} \\
		\cline{2-7}
		  & \textbf{P} &  \textbf{R}    & \textbf{F1} & \textbf{P} &  \textbf{R}    & \textbf{F1}     \\
		\hline
		\small{RGB + L1}    & 67.1\% & 68.3\% & 67.7\% & \textbf{80.1\%} & 77.5\%  & 78.7\% \\
		\small{RGB + L2}    & 49.7\% & 47.4\% & 48.5\% & 64.8\% & 74.0\%  & 69.1\% \\
		\small{RGB + SSIM}  & 62.0\% & 70.4\% & 65.9\% & 78.5\% & \textbf{84.7\%} & 81.5\% \\
		\small{YUV + L1}    & \textbf{67.4\%} & 68.3\% & \textbf{67.9\%} & 80.0\% & 77.8\%  & 78.9\% \\
		\small{YUV + L2}    & 50.9\% & 51.3\% & 51.1\% & 68.3\% & 68.2\%  & 68.2\% \\
		\small{YUV + SSIM}  & 62.2\% & 69.6\% & 65.7\% & 78.5\% & 83.5\%  & 80.9\% \\
		\small{Hist + L1}   & 61.2\% & 66.2\% & 63.6\% & 79.8\% & 76.1\%  & 77.9\% \\
		\small{Hist + L2}   & 62.5\% & 65.2\% & 63.8\% & 79.0\% & 72.3\%  & 75.5\%  \\
		\small{Hist + SSIM} & 51.4\% & 58.7\% & 54.8\% & 70.0\% & 52.2\%  & 59.8\% \\
		\small{HOG + L1}    & 57.6\% & 71.1\% & 63.6\% & 76.6\% & 83.1\%  & 79.7\% \\
		\small{HOG + L2}    & 56.5\% & 56.4\% & 56.4\% & 73.8\% & 83.3\%  & 78.2\% \\
		\small{HOG + SSIM}  & 61.0\% & \textbf{72.0\%} & 66.0\% & 79.2\% & \textbf{84.7\%}  & \textbf{81.9\%} \\
		\small{SIFT}        & 52.5\% & 58.4\% & 55.3\% & 70.3\% & 71.6\%  & 71.0\% \\
		\hline
	\end{tabular}
	\label{tab:action_segmentation_results_manual}
\end{table}

\cref{tab:action_segmentation_results_manual} shows the performance of the video segmentation phase using different features and feature distance functions.
Among all combinations, YUV+L1 performs the best (67.9\% F1 score) on Android recordings while HOG+SSIM (81.9\% F1 score) performs the best on iOS recordings.

Among all features, color histogram performed relatively the worst as it simply aggregates the general image features and somewhat downplays the salient changes.
RGB and YUV features performed similarly as they essentially describe the same features with different representations.
The HOG feature achieved the best recall (72.0\% in Android recordings, and 84.7\% in iOS recordings), which suggests that it effectively captures UI changes.

Among all feature distance functions, L2 distance had the worst performance, as it may overemphasize large changes. However, a distinguishable change does not necessarily imply a change in UI states. For example, changes in advertisement banner should not be considered as a new UI state.
SSIM had the best performance, as it is a perceptual metric, which is able to capture general information about the image.

Overall, our method can effectively segment app usage videos into interaction clips when UI states have salient differences. We would like to share insights when our method fails to predict a keyframe.
We missed keyframes when user interactions lead to a subtle change (or no change) on the UI. For example, when users select an item in a list, a small checkmark will appear. Such a small difference may be ignored by our simple feature extraction methods. Understanding the whole screen context would help us capture this important change on the UI. The effect of these errors are they reduce recall.
We predicted extra keyframes on animations that are not caused by user interactions. For example, when users enter an image-heavy screen, a loading animation may appear while waiting. Similarly, when users download a file, a progress bar updates frequently and may automatically move to the next UI state once the file is downloaded. In the future, we should recognize these common animations. The effect of these errors reduce precision.

We also obtained insights from the performance differences we found between iOS and Android recordings. Our method may predict frames with changing advertisements as extra keyframes, which are not caused by user interactions. From our observations, Android apps tended to contain more banner advertisements while iOS apps displayed fewer advertisements. In addition, the Android emulator we used may have higher latency than their physical counterpart devices. Because the emulator takes longer for UI rendering and UI transitions, this makes it harder to distinguish UI rendering and transitions from user interactions.

\begin{table}[t]
	\centering
	\caption{Experimental results for interaction classification on recordings from top-downloaded apps on Android and iOS, including Precision (P), Recall (R), and F1 Score}
	\begin{tabular}{| l | c c c | c c c |}
		
		\hline
		& \multicolumn{3}{c|}{\textbf{Android}} & \multicolumn{3}{c|}{\textbf{iOS}} \\
		\cline{2-7}
		  & \textbf{P} &  \textbf{R}    & \textbf{F1} & \textbf{P} &  \textbf{R}    & \textbf{F1}     \\
		\hline
		\small{Tap}             & 94.6\% & 91.9\%  & 92.7\% & 94.6\%  & 89.2\%  & 91.8\%  \\
		\small{Type}            & 74.4\% & 87.9\%  & 80.6\% & 57.1\%  & 100.0\% & 72.7\%  \\
		\small{Swipe Up}        & 92.9\% & 81.2\%  & 86.7\% & 90.2\%  & 74.3\%  & 81.5\%  \\
		\small{Swipe Down}      & 46.7\% & 53.8\%  & 50.0\% & 64.3\%  & 75.0\%  & 69.2\%  \\
		\small{Swipe Left}      & 46.2\% & 100.0\% & 63.2\% & 33.3\%  & 100.0\% & 50.0\%  \\
		\small{Swipe Right}     & 75.0\% & 100.0\% & 85.7\% & 100.0\% & 100.0\% & 100.0\% \\
		\hline
		\small{Macro Avg.}      & 71.4\% & 85.8\% & 76.5\% & 73.2\% & 89.8\% & 77.5\% \\
		\small{Weighted Avg.}   & 90.4\% & 89.3\% & 89.6\% & 90.3\% & 87.6\% & 88.3\% \\
		\hline
		        & \multicolumn{3}{c|}{Accuracy: 89.3\%} & \multicolumn{3}{c|}{Accuracy: 87.6\%} \\
		\hline
	\end{tabular}
	\label{tab:action_classifictaion_results_manual}
\end{table}

\subsection{Phase 2---Interaction Classification}
\label{sec:evaluation:interaction_classification}

We evaluated our model’s performance in interaction classification on usage recordings from top-downloaded iOS and Android apps. 

\cref{tab:action_classifictaion_results_manual} shows the performance of the interaction classification phase, which performs well on both iOS and Android app recordings (\revise{87.6\%} and \revise{89.3\%} accuracy respectively). Our method achieves high recall in most interaction types (except \textit{swipe down} on Android), and gets high F1 scores in \textit{tap}, \textit{swipe up}, and \textit{swipe right}. 

\textit{Swipe left} and \textit{swipe down} had the worst performance.
As they have only 9 and 27 samples out of 1,052 interactions, their precisions can be impacted by incorrect predictions from the other interaction types. Here is an example of failures of \textit{swipe left}: the screen scrolls horizontally when users \textit{tap} on the next segmented control, which has the same visual effect of \textit{swipe left}. We also found that half of \textit{swipe down} interactions on Android were recognized incorrectly as \textit{tap}. Most of these failures were related to a date/time picker: text in the pickers are smaller and faded to highlight currently selected text, which reduced accuracy of OCR that our heuristics rely on. Some failures in \textit{swipe down} happen when users \textit{tap} a button in the bottom actionsheet; the actionsheet moves down and disappears, creating a similar visual effect of \textit{swipe down}. Not surprisingly, we also found that most false positives are from \textit{tap}---the majority of interactions in our dataset.

Our method had reasonable performance on \textit{type} interactions. However, when the virtual keyboard appears, users may still perform non-type interactions like \textit{tap}. In the future, we should focus on the visual changes inside the virtual keyboard to confirm \textit{type} interactions.

\subsection{Phase 3---Interaction Localization}
\label{sec:evaluation:interaction_localization}

We evaluated our model’s performance in interaction localization on the large-scale Rico dataset, and usage recordings from top-downloaded iOS and Android apps.

\cref{tab:action_localization_results} shows the interaction localization performance of our system, and several baseline methods as comparison.
The first baseline is \textbf{Humanoid}~\cite{li2019humanoid}, which predicts the next interaction given the previous three UI screens. It follows a RNN-style method to encode frame features step-by-step, and predicts the heatmap of possible interaction points using a decoder module.
There are also variants of our model as baselines: \textbf{HM2D} (shorts for heatmap 2D) directly uses 2D convolutional layers to learn the semantics from animations, while \textbf{HM3D} instead uses 3D convolutional layers to learn the temporal and spatial features from animations through several layers.
The default \textbf{HM3D} contains a UNet-style shortcut, which is expected to help the model refine the high-level abstract features using the features from shallow layers.
We also consider a variant of \textbf{HM3D without shortcut} to see the impact from the shortcut module.
Another variant close to our model is \textbf{HM3D + 2D}, while our final system (\textbf{HM3D + 2D + Heuristics}) includes heuristics to improve performance. 
We also compared the performance when the input contained 8 or 16 frames from interaction clips.

\begin{table}[t]
  \caption{Accuracy of our interaction localization model compared with several baselines for the Rico-Test dataset and our manually collected recordings.
  }
  \label{tab:action_localization_results}
  \resizebox{0.5\textwidth}{!}{
  \begin{tabular}{|l | c | c |c |c|}
    \hline
    \multirow{2}{*}{} & \multirow{2}{*}{ \textbf{\#{Frames}}} & \multirow{2}{*}{\textbf{Rico-Test}} & \multicolumn{2}{c|}{\textbf{Recordings}} \\
    \cline{4-5}
    & & & \textbf{Android} & \textbf{iOS} \\
    \hline
    \texttt{Humanoid}               & 8/16 & 29.5\%/29.5\%                   & 8.8\%/8.8\%                         & 9.7\%/8.7\%      \\
    \texttt{HM2D}                   & 8/16 & 61.8\%/52.4\%                   & 34.3\%/21.7\%                       & 28.1\%/21.1\%    \\
    \texttt{HM3D w/o shortcut}      & 8/16 & 66.7\%/66.9\%                   & 52.8\%/53.0\%                       & 36.2\%/38.9\%    \\
    \texttt{HM3D}       & 8/16 & 67.9\%/67.0\%                   & 53.0\%/54.3\%                       & 40.4\%/38.5\%    \\
    \texttt{HM3D+2D}              & 8/16 & \textbf{69.1\%}/\textbf{67.9\%} & 52.3\%/54.3\%                       & 41.0\%/40.0\%    \\
    \texttt{HM3D+2D+Heuristics} & 8/16 & \textbf{69.1\%}/\textbf{67.9\%} & \textbf{53.6\%}/\textbf{56.2\%}     & \textbf{41.4\%}/\textbf{40.4\%}  \\
    \hline
  \end{tabular}
  }
\end{table}

\begin{table*}
  \caption{The recall of our interaction localization model as compared with several baselines,  across 6 common UI types in the Rico Dataset. \revise{The table shows the results for models trained on 8 frames / 16 frames.}
  }
  \label{tab:localization_ele_type}
  \begin{tabular}{|l | c | c |c |c | c | c | c |}
    \hline
                                     & \textbf{Text} & \textbf{Button} & \textbf{ImageView}  & \textbf{ImageButton}   & \textbf{System Nav. Bar}  & \textbf{Others}        \\
    \hline
    \texttt{Number}                  & \textit{783}  & \textit{625}    & \textit{521}  & \textit{531}  & \textit{388}    & \textit{1,592}         \\
    \hline
    \texttt{Humanoid}                & 1.1\%/1.3\%   & 3.5\%/3.7\%   & 17.1\%/17.1\%  & 68.7\%/68.7\%  & N/A      & 51.6\%/51.6\%    \\
    \texttt{HM2D}                    & 33.3\%/20.1\% & 63.8\%/51.7\% & 44.7\%/34.5\%  & 79.3\%/77.4\%  & 82.0\%/62.6\%    & 69.5\%/62.8\%    \\
    \texttt{HM3D w/o shortcut}       & 43.0\%/43.3\% & 68.6\%/66.7\% & 45.9\%/46.8\%  & 83.4\%/84.4\%  & 87.4\%/90.2\%    & 72.7\%/72.5\%   \\
    \texttt{HM3D}        & 44.3\%/40.2\% & 70.4\%/67.4\% & 48.9\%/49.5\%  & 83.1\%/85.3\%  & 89.2\%/89.9\%    & 74.1\%/73.3\%   \\
    \texttt{HM3D + 2D} & 44.8\%/43.3\% & 70.9\%/69.9\% & 49.1\%/47.4\%  & 85.5\%/86.1\%  & 89.4\%/89.9\%    & 75.4\%/73.6\%  \\
    \hline
  \end{tabular}
\end{table*}

\subsubsection{Performance on Rico Dataset}
Our system outperformed all baselines, reaching 69.1\% accuracy when the input contains 8 frames from interaction clip. We found the 3D convolutional network (HM3D) better captured the temporal features from interactions than RNN-style (Humanoid) and 2D convolutional based network (HM2D). RNN-style model encodes each frame, and the compressed frame may lose information before it is fed into the next RNN cell. A 2D-style model heavily relies on the first layer to capture the temporal features among frames, while a 3D-style model gradually learns the semantics of animations through several layers. UNet-style shortcut and additional 2D modules both help our model to better learn features and slightly boosted the performance. Applying heuristics also improved model performance in recent app recordings.

We then analyzed the performance across UI element types as they have different visual effects. We considered six common UI types in Rico dataset, namely: ImageButton, ImageView, TextView, Button, and System Bottom Navigation Bar. The rest of UI elements fall into ``Other'' type.
From \cref{tab:localization_ele_type}, we found TextView and ImageView elements had worse performance compared to other UI types. Other UI types are tappable by default and they have animations provided by the system UI framework. TextView and ImageView are not tappable unless developers specify the property or create a customized event listener. Therefore, these two UI types are more likely to have a special animation effect or no visual effect.
In contrast, all models performed best on the system back button (except for Humanoid) and ImageButton, which almost always provide visual feedback when users \textit{tap} them.

\subsubsection{Performance on Recent iOS and Android App Recordings} Our model is trained on the Rico dataset collected 4 years ago, and we wanted to investigate the feasibility of our method on recent mobile apps. As shown in \cref{tab:action_localization_results}, we found that the performance of all models degrade.

Since the release of the Rico dataset, design principles and UI styles have changed substantially in recent mobile apps.
One example is the redesign of system bottom navigation bar. Previously, Android apps avoided using the tab bar at the bottom of the screen (side menu drawer is a replacement), because users could easily tap system bottom navigation bar by mistake. Nowadays, the system bottom navigation bar no longer shows buttons, but only a subtle bar with space to enable gesture navigation. Android apps are more likely to use the tab bar instead of the menu drawer.
From the Rico dataset, we randomly selected 100 apps and sampled one interaction from each app. Only 6\% of apps contain a bottom tab bar, while most of the recent iOS (87.5\%) and Android (75\%) apps we collected contain a bottom tab bar.

We also found the animation visual feedback to be more subdued now. For example, the text color may only slightly change after a \textit{tap}. After examining the failure cases in recent app recordings, we found that our model worked best when the animation visual feedback is more apparent. As seen in the first two keyframes of \cref{fig:segmentation}, users \textit{tap} on the bottom tab bar---which leads to a subtle change in the text color of a tab button and an obvious change in the main content. In the future, the understanding of all UIs on a screen~\cite{zhang2021screen} will help our model focus on important UI changes---for example, to prioritize tab button changes when a tab bar is detected.

Finally, we noticed a large performance discrepancy between iOS and Android recordings, as the differences in their designs are even more substantial than the differences between Rico and recent Android apps. A larger-scale app usage recording dataset in both iOS and Android, like Rico, will help our model better capture the interactions under these new UI paradigms.

\subsection{Phase 4---Interaction Replay}
\label{sec:evaluation:interaction_replay}

We evaluated our model's performance in interaction replay on usage recordings from top-downloaded iOS and Android apps. In order to evaluate the success rate of interaction replay, the first two authors also collected the same app interaction traces on devices with different resolutions (Pixel 4 XL running Android 11 and iPhone 11 Pro Max running iOS 15). It is a manual replay process that will not stop by error in one step.

To focus on the performance of replay module itself, we directly used the annotated interactions as a ground truth to avoid the errors that propagate from each step during interaction extraction.

During this new collection, we took notes of four problems that prevented us from replaying 23 interactions on the target devices.
First, some pop-ups windows appear occasionally while replaying and required us to perform extra steps to close them.
These pop-up windows include advertisements, instruction hints, rating requests, and permission requests.
Second, some interactions were related to specific time that were no longer available.
For example, when we try to replay the interaction in a different month, the option of previous month may no longer exist and thus we cannot replay the exact same interaction.
Third, apps may contain dynamic content that changes the required user interactions.
For example, in a recording, we need to \textit{swipe up} five times to reach the target element to \textit{tap}; in the updated content, we only need to \textit{swipe up} twice.
Fourth, apps add and remove features in updates.
Such updates could lead to changes in the UI layout and UI transitions, removing an existing UI element, or affecting the navigation logic.
We counted the cases of each problem during the replay process.
Among failure cases in iOS | Android recordings, 9 | 12 are relevant to the pop-up windows, 1 | 1 are relevant to the time, 6 | 5 are relevant to dynamic content, and 3 | 5 are relevant to app updates.

For each interaction, our system found a matching target UI element on newly collected screens. The first two authors manually examined the matching result to determine whether interactions on the matched UI could lead to the expected next UI state. The majority of interactions could be correctly replayed in iOS and Android apps (84.1\% and 78.4\% respectively).

\begin{figure*}
	\centering
	\includegraphics[width=0.84\textwidth]{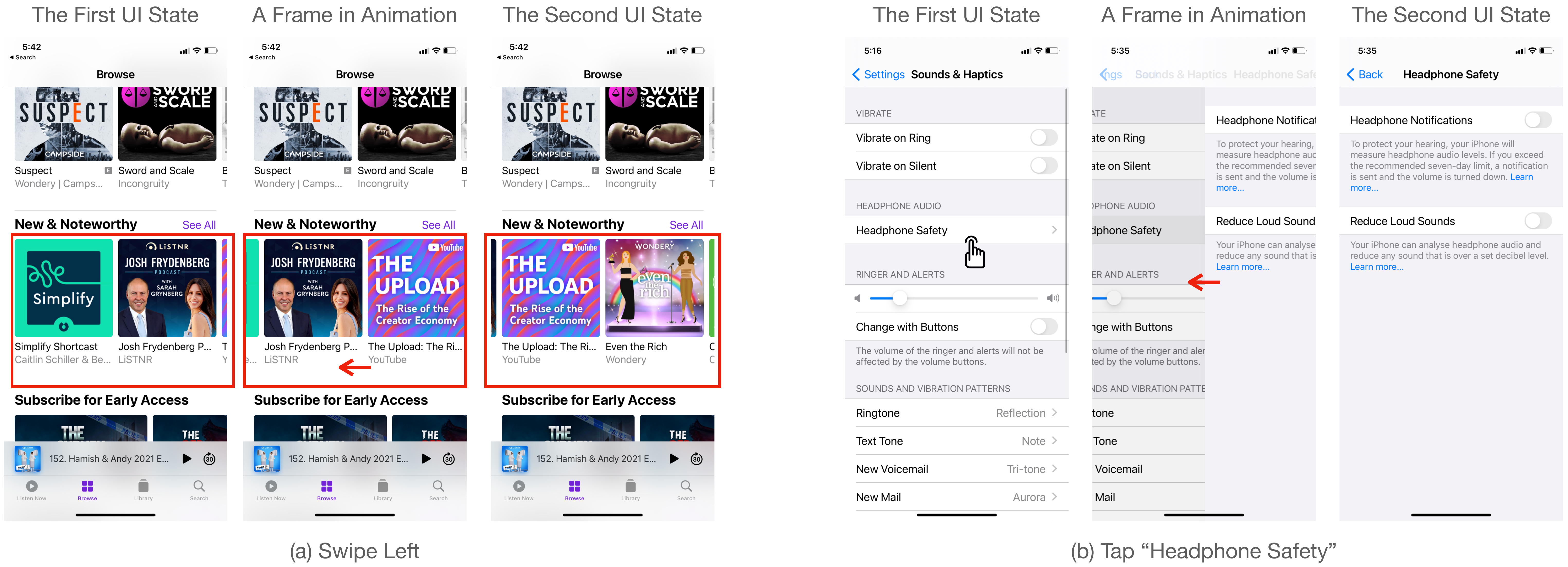}
	\caption{Examples of patterns that our interaction classification heuristics examine to classify a \textit{swipe left} interaction, including (a) a \textit{swipe left} interaction will likely change at least 3 text elements, and will not change the UI title, and (b) a \textit{tap} interaction instead is likely to change the title.}
	\label{fig:swipe_left_vs_tap}
	\Description{The figure shows two examples of interactions: \textit{swipe right} and \textit{tap}. On the left, we show three screenshots showing the transition between a first UI state and a new UI state after swiping left where a middle section has been scrolled. On the right, we show three screenshots to visualize a \textit{tap} interaction on setting "Headphone Safety". The UI in the first screenshot has the title "Sounds & Haptics", the UI in the second screenshot is in transition, and the UI in the third screenshot has the title "Headphone Safety"}
\end{figure*}

Here are some common failure cases in UI element matching: Image content may change across different sessions or change due to personalization (e.g., albums in \cref{fig:swipe_left_vs_tap}(a) are different in different accounts). Therefore, our image template matching would not find the same image. There can also be multiple UI elements contain the same text, and our text matching may find the wrong target text element. Beyond the scope of this paper, a deeper understanding of UI will help in resolving these failure cases: after our system learns what content is dynamic and what content is repetitive, it can replay the interaction on the UI element that has the same relative position in the UI structure.

\section{Discussion}

\revise{
\textbf{Datasets.}
We evaluated our proposed system with two different datasets. First, we use the Rico dataset to evaluate our localization models in a large-scale experiment, even though the Rico dataset only contains interactions with Android apps. The addition of a large-scale interaction dataset of iOS apps would help better illustrate the advantages and disadvantages of our system. To mitigate this issue, we then collected interaction traces from two top-downloaded apps from each app category that were available on both the iOS and Android platforms. We used these traces to better evaluate the generalizability and robustness of the proposed system. To ensure the manual recordings were representative, all authors together discussed and selected the tasks to be performed, ensuring that they covered the key features of all apps based on their descriptions in the app stores.  While each type of interaction has a different number of trials, we believe the diversity and representativenss of the collected apps and interactions mitigates some of the potential issues. We also report the detailed results for each interaction to better illustrate the performance on different interactions. Our future work will include more data to better examine and improve our system. Moreover, in the future we will examine extending our work to other input devices of different resolutions, such as larger-screened tablets. 
}

\textbf{Opportunities for performance improvements.} Our approach has several opportunities for performance improvements. First, several limitations in detecting interactions were a result of modern UI interaction paradigms which were not available when the Rico dataset was released. Consequently, we expect that a large-scale dataset of \emph{recent} apps on multiple mobile platforms would improve our system performance. Such a dataset would provide relevant data to train machine learning models to enhance our heuristic-based video segmentation and interaction classification phases, as well as and make our interaction localization model more generalizable to apps on iOS and Android platforms.

Our current localization model relied on video frames and their corresponding pixel-data, achieving around 70\% accuracy. This localization model could be improved in several ways. First, we could consider running UI detection on a screen to predict the interactable UI elements, as in \citet{zhang2021screen, chen2020object}. These UIs have higher probabilities for \textit{tap} interactions. Additional labels~\cite{chen2020unblind, chen2022towards} for some image-based interactable elements can also improve the system. Second, we tried a simple heuristic-based method to identify the connections between \textit{tapped} text and the title in new UI state. Deeper understanding of the content of text elements would support inferring the interaction between the two UI states. For example, ActionBert~\cite{he2020actionbert} demonstrates that it is possible to predict connection elements between two UI states---even without leveraging animation.

We proposed a straightforward method to replay interactions. In our evaluation, it works well on the same (or similar) app versions for a different device. This assumption applies to some applications---for example, automated or regression testing for the the same app version---but other scenarios, such as making app tutorials, may require using multiple app versions. Another challenge is to replay the interactions for the same app in different languages, which would enable cross-locale applications. Collecting the interaction traces in different settings (e.g., app versions, languages) would provide more signals for our interaction replay phase. Our current system only tests recordings within one app, while many tasks involve multiple apps. Learning the transition between apps will enable cross-app interaction extraction and replay to complete more complicated tasks.

We think of our pixel-based approach as a general technique that is also in some ways a \emph{lower-bound} on accuracy: by design, it does not take advantage of additional metadata that could potentially further improve its performance. Incorporating metadata, if available---such as the UI framework used within the app, platform, and version---could boost the performance of our approach.  Of course, the disadvantage of metadata is that it is not always available, difficult to extract, or unnecessarily constraints and couples the model to the metadata.

\textbf{Applications of extracting replayable interactions.} There are several applications that may benefit from our methods to extract and replay user interactions from video pixels. A straightforward application is to allow users to \emph{annotate interactions on existing videos}. For example, they may have a screen recording and have difficulty figuring out how the user in the screen recording is getting to a particular screen. Our system could be used to provide on-screen annotations of our inferred interactions as the user plays the video. 

For \emph{app bug reports}, users or QA testers could create videos of issues in apps, which developers could then replay within their own development environment to reproduce. Similarly, end-users could upload videos to demonstrate app usage problems:  automatically identifying the interactions in these videos could minimize or eliminate the errors introduced by more manual identification procedures. In \emph{automated app testing}, QA testers can sometimes only run apps on unmodified devices, which do not allow special recording tools or collection of metadata. Our method extracts interaction traces from app usage videos, and then replays them on other devices to test. After collecting a larger-scale app dataset on multiple platforms, our pixel-based method could potentially enable cross-platform testing without relying on the platform-specific testing APIs. 

Finally, our approach could be used in \emph{app tutorials}. As one example, \revise{people with limited mobile usage experience} and people with cognitive impairments sometimes require help from others (such as their caregivers) to use a new app or an updated version of app. Our method might be applied to automatically create app tutorials from app usage video recorded by users who better understand and can demonstrate the app functionality. Then, people in need can replay interactions on their own mobile devices, or learn how to use apps with an on-device, interactive tutorials~\cite{harms2011improving}.

\section{Conclusion}
In this paper, we introduced a novel approach to automatically extract and replay interactions from video pixels without requiring additional settings, recording tools, or source code access. Our approach automatically segments interactions from a video, classifies interaction types, and locates target UI elements for replay. We trained our system using the large-scale Rico dataset for Android, evaluated its effectiveness, and demonstrated the feasibility of learning interaction locations for recent iOS and Android apps. Our prototype can successfully replay the majority of the interactions. The results of this work suggest that extracting replayable interactions is a useful mechanism that potentially benefits a variety of different applications and scenarios.

\bibliographystyle{ACM-Reference-Format}
\bibliography{reference}


\begin{thebibliography}{33}


\ifx \showCODEN    \undefined \def \showCODEN     #1{\unskip}     \fi
\ifx \showDOI      \undefined \def \showDOI       #1{#1}\fi
\ifx \showISBNx    \undefined \def \showISBNx     #1{\unskip}     \fi
\ifx \showISBNxiii \undefined \def \showISBNxiii  #1{\unskip}     \fi
\ifx \showISSN     \undefined \def \showISSN      #1{\unskip}     \fi
\ifx \showLCCN     \undefined \def \showLCCN      #1{\unskip}     \fi
\ifx \shownote     \undefined \def \shownote      #1{#1}          \fi
\ifx \showarticletitle \undefined \def \showarticletitle #1{#1}   \fi
\ifx \showURL      \undefined \def \showURL       {\relax}        \fi
\providecommand\bibfield[2]{#2}
\providecommand\bibinfo[2]{#2}
\providecommand\natexlab[1]{#1}
\providecommand\showeprint[2][]{arXiv:#2}

\bibitem[\protect\citeauthoryear{Bao, Li, Xing, Wang, Xia, and Zhou}{Bao
  et~al\mbox{.}}{2017}]%
        {bao2017extracting}
\bibfield{author}{\bibinfo{person}{Lingfeng Bao}, \bibinfo{person}{Jing Li},
  \bibinfo{person}{Zhenchang Xing}, \bibinfo{person}{Xinyu Wang},
  \bibinfo{person}{Xin Xia}, {and} \bibinfo{person}{Bo Zhou}.}
  \bibinfo{year}{2017}\natexlab{}.
\newblock \showarticletitle{Extracting and analyzing time-series HCI data from
  screen-captured task videos}.
\newblock \bibinfo{journal}{\emph{Empirical Software Engineering}}
  \bibinfo{volume}{22}, \bibinfo{number}{1} (\bibinfo{year}{2017}),
  \bibinfo{pages}{134--174}.
\newblock


\bibitem[\protect\citeauthoryear{Bernal-C{\'a}rdenas, Cooper, Moran, Chaparro,
  Marcus, and Poshyvanyk}{Bernal-C{\'a}rdenas et~al\mbox{.}}{2020}]%
        {bernal2020translating}
\bibfield{author}{\bibinfo{person}{Carlos Bernal-C{\'a}rdenas},
  \bibinfo{person}{Nathan Cooper}, \bibinfo{person}{Kevin Moran},
  \bibinfo{person}{Oscar Chaparro}, \bibinfo{person}{Andrian Marcus}, {and}
  \bibinfo{person}{Denys Poshyvanyk}.} \bibinfo{year}{2020}\natexlab{}.
\newblock \showarticletitle{Translating video recordings of mobile app usages
  into replayable scenarios}. In \bibinfo{booktitle}{\emph{Proceedings of the
  ACM/IEEE 42nd International Conference on Software Engineering}}.
  \bibinfo{pages}{309--321}.
\newblock


\bibitem[\protect\citeauthoryear{Chen, Chen, Xing, Xu, Zhut, Li, and Wang}{Chen
  et~al\mbox{.}}{2020a}]%
        {chen2020unblind}
\bibfield{author}{\bibinfo{person}{Jieshan Chen}, \bibinfo{person}{Chunyang
  Chen}, \bibinfo{person}{Zhenchang Xing}, \bibinfo{person}{Xiwei Xu},
  \bibinfo{person}{Liming Zhut}, \bibinfo{person}{Guoqiang Li}, {and}
  \bibinfo{person}{Jinshui Wang}.} \bibinfo{year}{2020}\natexlab{a}.
\newblock \showarticletitle{Unblind your apps: Predicting natural-language
  labels for mobile gui components by deep learning}. In
  \bibinfo{booktitle}{\emph{2020 IEEE/ACM 42nd International Conference on
  Software Engineering (ICSE)}}. IEEE, \bibinfo{pages}{322--334}.
\newblock


\bibitem[\protect\citeauthoryear{Chen, Swearngin, Wu, Barik, Nichols, and
  Zhang}{Chen et~al\mbox{.}}{2022}]%
        {chen2022towards}
\bibfield{author}{\bibinfo{person}{Jieshan Chen}, \bibinfo{person}{Amanda
  Swearngin}, \bibinfo{person}{Jason Wu}, \bibinfo{person}{Titus Barik},
  \bibinfo{person}{Jeffrey Nichols}, {and} \bibinfo{person}{Xiaoyi Zhang}.}
  \bibinfo{year}{2022}\natexlab{}.
\newblock \showarticletitle{Towards Complete Icon Labeling in Mobile
  Applications}. In \bibinfo{booktitle}{\emph{CHI Conference on Human Factors
  in Computing Systems}} (New Orleans, LA, USA) \emph{(\bibinfo{series}{CHI
  '22})}. \bibinfo{publisher}{Association for Computing Machinery},
  \bibinfo{address}{New York, NY, USA}, Article \bibinfo{articleno}{387},
  \bibinfo{numpages}{14}~pages.
\newblock
\showISBNx{9781450391573}
\urldef\tempurl%
\url{https://doi.org/10.1145/3491102.3502073}
\showDOI{\tempurl}


\bibitem[\protect\citeauthoryear{Chen, Xie, Xing, Chen, Xu, Zhu, and Li}{Chen
  et~al\mbox{.}}{2020b}]%
        {chen2020object}
\bibfield{author}{\bibinfo{person}{Jieshan Chen}, \bibinfo{person}{Mulong Xie},
  \bibinfo{person}{Zhenchang Xing}, \bibinfo{person}{Chunyang Chen},
  \bibinfo{person}{Xiwei Xu}, \bibinfo{person}{Liming Zhu}, {and}
  \bibinfo{person}{Guoqiang Li}.} \bibinfo{year}{2020}\natexlab{b}.
\newblock \showarticletitle{Object detection for graphical user interface: old
  fashioned or deep learning or a combination?}. In
  \bibinfo{booktitle}{\emph{proceedings of the 28th ACM joint meeting on
  European Software Engineering Conference and Symposium on the Foundations of
  Software Engineering}}. \bibinfo{pages}{1202--1214}.
\newblock


\bibitem[\protect\citeauthoryear{Dalal and Triggs}{Dalal and Triggs}{2005}]%
        {dalal2005histograms}
\bibfield{author}{\bibinfo{person}{Navneet Dalal} {and} \bibinfo{person}{Bill
  Triggs}.} \bibinfo{year}{2005}\natexlab{}.
\newblock \showarticletitle{Histograms of oriented gradients for human
  detection}. In \bibinfo{booktitle}{\emph{2005 IEEE computer society
  conference on computer vision and pattern recognition (CVPR'05)}},
  Vol.~\bibinfo{volume}{1}. Ieee, \bibinfo{pages}{886--893}.
\newblock


\bibitem[\protect\citeauthoryear{Deka, Huang, Franzen, Hibschman, Afergan, Li,
  Nichols, and Kumar}{Deka et~al\mbox{.}}{2017}]%
        {deka2017rico}
\bibfield{author}{\bibinfo{person}{Biplab Deka}, \bibinfo{person}{Zifeng
  Huang}, \bibinfo{person}{Chad Franzen}, \bibinfo{person}{Joshua Hibschman},
  \bibinfo{person}{Daniel Afergan}, \bibinfo{person}{Yang Li},
  \bibinfo{person}{Jeffrey Nichols}, {and} \bibinfo{person}{Ranjitha Kumar}.}
  \bibinfo{year}{2017}\natexlab{}.
\newblock \showarticletitle{Rico: A mobile app dataset for building data-driven
  design applications}. In \bibinfo{booktitle}{\emph{Proceedings of the 30th
  Annual ACM Symposium on User Interface Software and Technology}}.
  \bibinfo{pages}{845--854}.
\newblock


\bibitem[\protect\citeauthoryear{Duan, Bai, Xie, Qi, Huang, and Tian}{Duan
  et~al\mbox{.}}{2019}]%
        {duan2019centernet}
\bibfield{author}{\bibinfo{person}{Kaiwen Duan}, \bibinfo{person}{Song Bai},
  \bibinfo{person}{Lingxi Xie}, \bibinfo{person}{Honggang Qi},
  \bibinfo{person}{Qingming Huang}, {and} \bibinfo{person}{Qi Tian}.}
  \bibinfo{year}{2019}\natexlab{}.
\newblock \showarticletitle{Centernet: Keypoint triplets for object detection}.
  In \bibinfo{booktitle}{\emph{Proceedings of the IEEE/CVF International
  Conference on Computer Vision}}. \bibinfo{pages}{6569--6578}.
\newblock


\bibitem[\protect\citeauthoryear{Feng and Chen}{Feng and Chen}{2022}]%
        {feng2022gifdroid}
\bibfield{author}{\bibinfo{person}{Sidong Feng} {and} \bibinfo{person}{Chunyang
  Chen}.} \bibinfo{year}{2022}\natexlab{}.
\newblock \showarticletitle{GIFdroid: Automated Replay of Visual Bug Reports
  for Android Apps}. In \bibinfo{booktitle}{\emph{2022 IEEE/ACM 44th
  International Conference on Software Engineering (ICSE)}}. ACM.
\newblock


\bibitem[\protect\citeauthoryear{Guo, Kong, Rivera, Xu, and Bigham}{Guo
  et~al\mbox{.}}{2019}]%
        {guo2019statelens}
\bibfield{author}{\bibinfo{person}{Anhong Guo}, \bibinfo{person}{Junhan Kong},
  \bibinfo{person}{Michael Rivera}, \bibinfo{person}{Frank~F Xu}, {and}
  \bibinfo{person}{Jeffrey~P Bigham}.} \bibinfo{year}{2019}\natexlab{}.
\newblock \showarticletitle{Statelens: A reverse engineering solution for
  making existing dynamic touchscreens accessible}. In
  \bibinfo{booktitle}{\emph{Proceedings of the 32nd Annual ACM Symposium on
  User Interface Software and Technology}}. \bibinfo{pages}{371--385}.
\newblock


\bibitem[\protect\citeauthoryear{Halpern, Zhu, Peri, and Reddi}{Halpern
  et~al\mbox{.}}{2015}]%
        {halpern2015mosaic}
\bibfield{author}{\bibinfo{person}{Matthew Halpern}, \bibinfo{person}{Yuhao
  Zhu}, \bibinfo{person}{Ramesh Peri}, {and} \bibinfo{person}{Vijay~Janapa
  Reddi}.} \bibinfo{year}{2015}\natexlab{}.
\newblock \showarticletitle{Mosaic: cross-platform user-interaction record and
  replay for the fragmented android ecosystem}. In
  \bibinfo{booktitle}{\emph{2015 IEEE International Symposium on Performance
  Analysis of Systems and Software (ISPASS)}}. IEEE, \bibinfo{pages}{215--224}.
\newblock


\bibitem[\protect\citeauthoryear{Harms, Kerr, and Kelleher}{Harms
  et~al\mbox{.}}{2011}]%
        {harms2011improving}
\bibfield{author}{\bibinfo{person}{Kyle~J Harms}, \bibinfo{person}{Jordana~H
  Kerr}, {and} \bibinfo{person}{Caitlin~L Kelleher}.}
  \bibinfo{year}{2011}\natexlab{}.
\newblock \showarticletitle{Improving learning transfer from stencils-based
  tutorials}. In \bibinfo{booktitle}{\emph{Proceedings of the 10th
  International Conference on Interaction Design and Children}}.
  \bibinfo{pages}{157--160}.
\newblock


\bibitem[\protect\citeauthoryear{He, Sunkara, Zang, Xu, Liu, Wichers,
  Schubiner, Lee, Chen, and y~Arcas}{He et~al\mbox{.}}{2020}]%
        {he2020actionbert}
\bibfield{author}{\bibinfo{person}{Zecheng He}, \bibinfo{person}{Srinivas
  Sunkara}, \bibinfo{person}{Xiaoxue Zang}, \bibinfo{person}{Ying Xu},
  \bibinfo{person}{Lijuan Liu}, \bibinfo{person}{Nevan Wichers},
  \bibinfo{person}{Gabriel Schubiner}, \bibinfo{person}{Ruby Lee},
  \bibinfo{person}{Jindong Chen}, {and} \bibinfo{person}{Blaise~Aguera y
  Arcas}.} \bibinfo{year}{2020}\natexlab{}.
\newblock \showarticletitle{ActionBert: Leveraging User Actions for Semantic
  Understanding of User Interfaces}.
\newblock \bibinfo{journal}{\emph{arXiv preprint arXiv:2012.12350}}
  (\bibinfo{year}{2020}).
\newblock


\bibitem[\protect\citeauthoryear{Kingma and Ba}{Kingma and Ba}{2014}]%
        {kingma2014adam}
\bibfield{author}{\bibinfo{person}{Diederik~P Kingma} {and}
  \bibinfo{person}{Jimmy Ba}.} \bibinfo{year}{2014}\natexlab{}.
\newblock \showarticletitle{Adam: A method for stochastic optimization}.
\newblock \bibinfo{journal}{\emph{arXiv preprint arXiv:1412.6980}}
  (\bibinfo{year}{2014}).
\newblock


\bibitem[\protect\citeauthoryear{Law and Deng}{Law and Deng}{2018}]%
        {law2018cornernet}
\bibfield{author}{\bibinfo{person}{Hei Law} {and} \bibinfo{person}{Jia Deng}.}
  \bibinfo{year}{2018}\natexlab{}.
\newblock \showarticletitle{Cornernet: Detecting objects as paired keypoints}.
  In \bibinfo{booktitle}{\emph{Proceedings of the European conference on
  computer vision (ECCV)}}. \bibinfo{pages}{734--750}.
\newblock


\bibitem[\protect\citeauthoryear{Li, Labutov, Li, Zhang, Shi, Ding, Mitchell,
  and Myers}{Li et~al\mbox{.}}{2018}]%
        {li2018appinite}
\bibfield{author}{\bibinfo{person}{Toby Jia-Jun Li}, \bibinfo{person}{Igor
  Labutov}, \bibinfo{person}{Xiaohan~Nancy Li}, \bibinfo{person}{Xiaoyi Zhang},
  \bibinfo{person}{Wenze Shi}, \bibinfo{person}{Wanling Ding},
  \bibinfo{person}{Tom~M Mitchell}, {and} \bibinfo{person}{Brad~A Myers}.}
  \bibinfo{year}{2018}\natexlab{}.
\newblock \showarticletitle{APPINITE: A Multi-Modal Interface for Specifying
  Data Descriptions in Programming by Demonstration Using Natural Language
  Instructions}. In \bibinfo{booktitle}{\emph{2018 IEEE Symposium on Visual
  Languages and Human-Centric Computing (VL/HCC)}}. IEEE,
  \bibinfo{pages}{105--114}.
\newblock


\bibitem[\protect\citeauthoryear{Li, Yang, Guo, and Chen}{Li
  et~al\mbox{.}}{2019}]%
        {li2019humanoid}
\bibfield{author}{\bibinfo{person}{Yuanchun Li}, \bibinfo{person}{Ziyue Yang},
  \bibinfo{person}{Yao Guo}, {and} \bibinfo{person}{Xiangqun Chen}.}
  \bibinfo{year}{2019}\natexlab{}.
\newblock \showarticletitle{Humanoid: A deep learning-based approach to
  automated black-box android app testing}. In \bibinfo{booktitle}{\emph{2019
  34th IEEE/ACM International Conference on Automated Software Engineering
  (ASE)}}. IEEE, \bibinfo{pages}{1070--1073}.
\newblock


\bibitem[\protect\citeauthoryear{Lin, Goyal, Girshick, He, and Doll{\'a}r}{Lin
  et~al\mbox{.}}{2017}]%
        {lin2017focal}
\bibfield{author}{\bibinfo{person}{Tsung-Yi Lin}, \bibinfo{person}{Priya
  Goyal}, \bibinfo{person}{Ross Girshick}, \bibinfo{person}{Kaiming He}, {and}
  \bibinfo{person}{Piotr Doll{\'a}r}.} \bibinfo{year}{2017}\natexlab{}.
\newblock \showarticletitle{Focal loss for dense object detection}. In
  \bibinfo{booktitle}{\emph{Proceedings of the IEEE international conference on
  computer vision}}. \bibinfo{pages}{2980--2988}.
\newblock


\bibitem[\protect\citeauthoryear{Nguyen and Liu}{Nguyen and Liu}{2015}]%
        {nguyen2015making}
\bibfield{author}{\bibinfo{person}{Cuong Nguyen} {and} \bibinfo{person}{Feng
  Liu}.} \bibinfo{year}{2015}\natexlab{}.
\newblock \showarticletitle{Making software tutorial video responsive}. In
  \bibinfo{booktitle}{\emph{Proceedings of the 33rd Annual ACM Conference on
  Human Factors in Computing Systems}}. \bibinfo{pages}{1565--1568}.
\newblock


\bibitem[\protect\citeauthoryear{Pfister, Charles, and Zisserman}{Pfister
  et~al\mbox{.}}{2015}]%
        {pfister2015flowing}
\bibfield{author}{\bibinfo{person}{Tomas Pfister}, \bibinfo{person}{James
  Charles}, {and} \bibinfo{person}{Andrew Zisserman}.}
  \bibinfo{year}{2015}\natexlab{}.
\newblock \showarticletitle{Flowing convnets for human pose estimation in
  videos}. In \bibinfo{booktitle}{\emph{Proceedings of the IEEE international
  conference on computer vision}}. \bibinfo{pages}{1913--1921}.
\newblock


\bibitem[\protect\citeauthoryear{Qian, Shang, Yan, Wang, and Chen}{Qian
  et~al\mbox{.}}{2020}]%
        {qian2020roscript}
\bibfield{author}{\bibinfo{person}{Ju Qian}, \bibinfo{person}{Zhengyu Shang},
  \bibinfo{person}{Shuoyan Yan}, \bibinfo{person}{Yan Wang}, {and}
  \bibinfo{person}{Lin Chen}.} \bibinfo{year}{2020}\natexlab{}.
\newblock \showarticletitle{Roscript: a visual script driven truly
  non-intrusive robotic testing system for touch screen applications}. In
  \bibinfo{booktitle}{\emph{Proceedings of the ACM/IEEE 42nd International
  Conference on Software Engineering}}. \bibinfo{pages}{297--308}.
\newblock


\bibitem[\protect\citeauthoryear{Ronneberger, Fischer, and Brox}{Ronneberger
  et~al\mbox{.}}{2015}]%
        {ronneberger2015u}
\bibfield{author}{\bibinfo{person}{Olaf Ronneberger}, \bibinfo{person}{Philipp
  Fischer}, {and} \bibinfo{person}{Thomas Brox}.}
  \bibinfo{year}{2015}\natexlab{}.
\newblock \showarticletitle{U-net: Convolutional networks for biomedical image
  segmentation}. In \bibinfo{booktitle}{\emph{International Conference on
  Medical image computing and computer-assisted intervention}}. Springer,
  \bibinfo{pages}{234--241}.
\newblock


\bibitem[\protect\citeauthoryear{seatgeek}{seatgeek}{2021}]%
        {fuzzywuzzy}
\bibfield{author}{\bibinfo{person}{seatgeek}.} \bibinfo{year}{2021}\natexlab{}.
\newblock \bibinfo{title}{GitHub - seatgeek/fuzzywuzzy}.
\newblock \bibinfo{howpublished}{\url{https://github.com/seatgeek/fuzzywuzzy}}.
\newblock
\newblock
\shownote{Accessed: 24/09/2021.}


\bibitem[\protect\citeauthoryear{Sprenkle, Gibson, Sampath, and
  Pollock}{Sprenkle et~al\mbox{.}}{2005}]%
        {sprenkle2005automated}
\bibfield{author}{\bibinfo{person}{Sara Sprenkle}, \bibinfo{person}{Emily
  Gibson}, \bibinfo{person}{Sreedevi Sampath}, {and} \bibinfo{person}{Lori
  Pollock}.} \bibinfo{year}{2005}\natexlab{}.
\newblock \showarticletitle{Automated replay and failure detection for web
  applications}. In \bibinfo{booktitle}{\emph{Proceedings of the 20th IEEE/ACM
  international conference on automated software engineering}}.
  \bibinfo{pages}{253--262}.
\newblock


\bibitem[\protect\citeauthoryear{Tran, Bourdev, Fergus, Torresani, and
  Paluri}{Tran et~al\mbox{.}}{2015}]%
        {tran2015learning}
\bibfield{author}{\bibinfo{person}{Du Tran}, \bibinfo{person}{Lubomir Bourdev},
  \bibinfo{person}{Rob Fergus}, \bibinfo{person}{Lorenzo Torresani}, {and}
  \bibinfo{person}{Manohar Paluri}.} \bibinfo{year}{2015}\natexlab{}.
\newblock \showarticletitle{Learning spatiotemporal features with 3d
  convolutional networks}. In \bibinfo{booktitle}{\emph{Proceedings of the IEEE
  international conference on computer vision}}. \bibinfo{pages}{4489--4497}.
\newblock


\bibitem[\protect\citeauthoryear{Tutorials}{Tutorials}{2021}]%
        {matchTemplate}
\bibfield{author}{\bibinfo{person}{OpenCV-Python Tutorials}.}
  \bibinfo{year}{2021}\natexlab{}.
\newblock \bibinfo{title}{Template Matching}.
\newblock
  \bibinfo{howpublished}{\url{https://opencv24-python-tutorials.readthedocs.io/en/stable/py_tutorials/py_imgproc/py_template_matching/py_template_matching.html}}.
\newblock
\newblock
\shownote{Accessed: 24/09/2021.}


\bibitem[\protect\citeauthoryear{tzutalin}{tzutalin}{2021}]%
        {labelImg}
\bibfield{author}{\bibinfo{person}{tzutalin}.} \bibinfo{year}{2021}\natexlab{}.
\newblock \bibinfo{title}{GitHub - tzutalin/labelImg}.
\newblock \bibinfo{howpublished}{\url{https://github.com/tzutalin/labelImg}}.
\newblock
\newblock
\shownote{Accessed: 24/09/2021.}


\bibitem[\protect\citeauthoryear{Wang, Bovik, Sheikh, and Simoncelli}{Wang
  et~al\mbox{.}}{2004}]%
        {wang2004image}
\bibfield{author}{\bibinfo{person}{Zhou Wang}, \bibinfo{person}{Alan~C Bovik},
  \bibinfo{person}{Hamid~R Sheikh}, {and} \bibinfo{person}{Eero~P Simoncelli}.}
  \bibinfo{year}{2004}\natexlab{}.
\newblock \showarticletitle{Image quality assessment: from error visibility to
  structural similarity}.
\newblock \bibinfo{journal}{\emph{IEEE transactions on image processing}}
  \bibinfo{volume}{13}, \bibinfo{number}{4} (\bibinfo{year}{2004}),
  \bibinfo{pages}{600--612}.
\newblock


\bibitem[\protect\citeauthoryear{Yu, Fang, Feng, Zhao, and Chen}{Yu
  et~al\mbox{.}}{2019}]%
        {yu2019lirat}
\bibfield{author}{\bibinfo{person}{Shengcheng Yu}, \bibinfo{person}{Chunrong
  Fang}, \bibinfo{person}{Yang Feng}, \bibinfo{person}{Wenyuan Zhao}, {and}
  \bibinfo{person}{Zhenyu Chen}.} \bibinfo{year}{2019}\natexlab{}.
\newblock \showarticletitle{Lirat: Layout and image recognition driving
  automated mobile testing of cross-platform}. In
  \bibinfo{booktitle}{\emph{2019 34th IEEE/ACM International Conference on
  Automated Software Engineering (ASE)}}. IEEE, \bibinfo{pages}{1066--1069}.
\newblock


\bibitem[\protect\citeauthoryear{Zhang, de~Greef, Swearngin, White, Murray, Yu,
  Shan, Nichols, Wu, Fleizach, et~al\mbox{.}}{Zhang et~al\mbox{.}}{2021}]%
        {zhang2021screen}
\bibfield{author}{\bibinfo{person}{Xiaoyi Zhang}, \bibinfo{person}{Lilian de
  Greef}, \bibinfo{person}{Amanda Swearngin}, \bibinfo{person}{Samuel White},
  \bibinfo{person}{Kyle Murray}, \bibinfo{person}{Lisa Yu}, \bibinfo{person}{Qi
  Shan}, \bibinfo{person}{Jeffrey Nichols}, \bibinfo{person}{Jason Wu},
  \bibinfo{person}{Chris Fleizach}, {et~al\mbox{.}}}
  \bibinfo{year}{2021}\natexlab{}.
\newblock \showarticletitle{Screen Recognition: Creating Accessibility Metadata
  for Mobile Applications from Pixels}. In
  \bibinfo{booktitle}{\emph{Proceedings of the 2021 CHI Conference on Human
  Factors in Computing Systems}}. \bibinfo{pages}{1--15}.
\newblock


\bibitem[\protect\citeauthoryear{Zhang, Ross, Caspi, Fogarty, and
  Wobbrock}{Zhang et~al\mbox{.}}{2017}]%
        {zhang2017interaction}
\bibfield{author}{\bibinfo{person}{Xiaoyi Zhang}, \bibinfo{person}{Anne~Spencer
  Ross}, \bibinfo{person}{Anat Caspi}, \bibinfo{person}{James Fogarty}, {and}
  \bibinfo{person}{Jacob~O Wobbrock}.} \bibinfo{year}{2017}\natexlab{}.
\newblock \showarticletitle{Interaction proxies for runtime repair and
  enhancement of mobile application accessibility}. In
  \bibinfo{booktitle}{\emph{Proceedings of the 2017 CHI conference on human
  factors in computing systems}}. \bibinfo{pages}{6024--6037}.
\newblock


\bibitem[\protect\citeauthoryear{Zhao, Xing, Chen, Xia, and Li}{Zhao
  et~al\mbox{.}}{2019}]%
        {zhao2019actionnet}
\bibfield{author}{\bibinfo{person}{Dehai Zhao}, \bibinfo{person}{Zhenchang
  Xing}, \bibinfo{person}{Chunyang Chen}, \bibinfo{person}{Xin Xia}, {and}
  \bibinfo{person}{Guoqiang Li}.} \bibinfo{year}{2019}\natexlab{}.
\newblock \showarticletitle{ActionNet: Vision-based workflow action recognition
  from programming screencasts}. In \bibinfo{booktitle}{\emph{2019 IEEE/ACM
  41st International Conference on Software Engineering (ICSE)}}. IEEE,
  \bibinfo{pages}{350--361}.
\newblock


\bibitem[\protect\citeauthoryear{Zhong, Zhang, and Chang}{Zhong
  et~al\mbox{.}}{1996}]%
        {zhong1996clustering}
\bibfield{author}{\bibinfo{person}{Di Zhong}, \bibinfo{person}{HongJiang
  Zhang}, {and} \bibinfo{person}{Shih-Fu Chang}.}
  \bibinfo{year}{1996}\natexlab{}.
\newblock \showarticletitle{Clustering methods for video browsing and
  annotation}. In \bibinfo{booktitle}{\emph{Storage and Retrieval for Still
  Image and Video Databases IV}}, Vol.~\bibinfo{volume}{2670}. International
  Society for Optics and Photonics, \bibinfo{pages}{239--246}.
\newblock


\end{thebibliography}
\pagebreak
\section{Appendix}

\begin{figure*}
	\centering
	\includegraphics[width=0.84\textwidth]{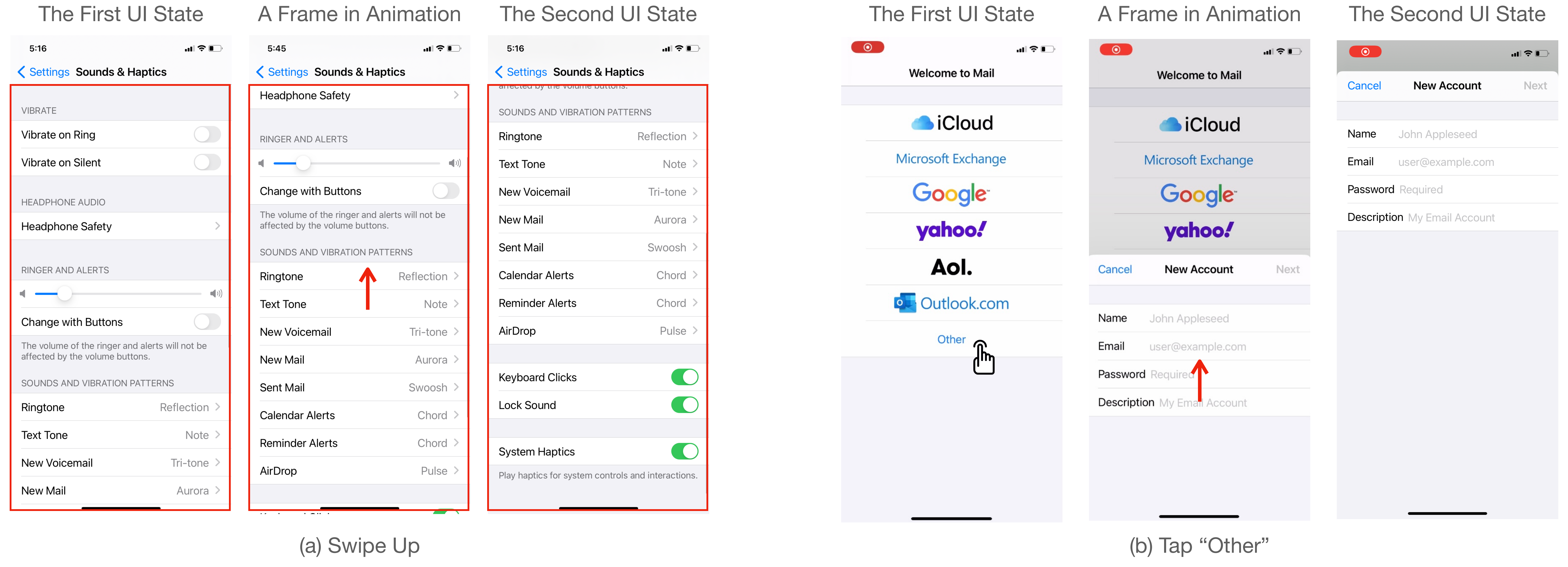}
 	\caption{Examples of patterns that our interaction classification heuristics examine to classify a \textit{swipe up} interaction, including (a) a \textit{swipe up} interaction will likely change several text elements, but will not change the UI title, and (b) \textit{tap} interaction instead is likely to change the title.}
	\label{fig:swipe_up_vs_tap}
	\Description{The figure shows two examples: a \textit{swipe up} interaction and a \textit{tap} interaction. On the left, we show three screenshots where the UI in each screenshot has the title "Sound & Haptics". The portion of the UI below the title has been scrolled slightly in each consecutive screenshot. On the right, we show three screenshots to visualize a \textit{tap} interaction. The UI in the first screenshot has the title "Welcome to Mail", the UI in the second screenshot shows the transition between the first and third UI states, and the UI in the third screneshot has the title "New Account" and a dialog has been opened.}
\end{figure*}

\end{document}